\documentclass[%
  superscriptaddress,
  twocolumn,
  10pt,
  aps,
  pre,
  nofootinbib, 
  final,letterpaper
]{revtex4-2}

\usepackage{graphicx}
\usepackage{hyperref}
\usepackage[utf8]{inputenc}
\usepackage{float}
\usepackage{amsmath,amsfonts,amssymb}

\begin{document}

\title{Effectiveness of Contact Tracing on Networks with Cliques}

\author{Abbas K. Rizi}
\affiliation{Department of Computer Science, School of Science, Aalto University, FI-00076, Finland}
\author{Leah A. Keating}
\affiliation{MACSI, Department of Mathematics and Statistics, University of Limerick, Limerick, V94 T9PX, Ireland}
\affiliation{Department of Mathematics, University of California, Los Angeles, CA 90095, USA}
\author{James P. Gleeson}
\affiliation{MACSI, Department of Mathematics and Statistics, University of Limerick, Limerick, V94 T9PX, Ireland}
\author{David J.P. O'Sullivan}
\affiliation{MACSI, Department of Mathematics and Statistics, University of Limerick, Limerick, V94 T9PX, Ireland}
\author{Mikko Kivelä}
\affiliation{Department of Computer Science, School of Science, Aalto University, FI-00076, Finland}

\date{\today}

\begin{abstract}
Contact tracing, the practice of isolating individuals who have been in contact with infected individuals, is an effective and practical way of containing disease spread. Here, we show that this strategy is particularly effective in the presence of social groups: Once the disease enters a group, contact tracing not only cuts direct infection paths but can also pre-emptively quarantine group members such that it will cut indirect spreading routes. We show these results by using a deliberately stylized model that allows us to isolate the effect of contact tracing within the clique structure of the network where the contagion is spreading. This will enable us to derive mean-field approximations and epidemic thresholds to demonstrate the efficiency of contact tracing in social networks with small groups. This analysis shows that contact tracing in networks with groups is more efficient the larger the groups are.
We show how these results can be understood by approximating the combination of disease spreading and contact tracing with a complex contagion process where every failed infection attempt will lead to a lower infection probability in the following attempts.
Our results illustrate how contact tracing in real-world settings can be more efficient than predicted by models that treat the system as fully mixed or the network structure as locally tree-like.
\end{abstract}

\maketitle

\section{Introduction}
\label{sec:intro}
 Contact tracing identifies, assesses, and manages people exposed to the disease through an infected individual \cite{brandt2022history}. This approach, inclusive of testing \cite{mercer2021testing, fogh2021testing} and isolating, has been a cornerstone in controlling disease spread and preventing outbreaks. The COVID-19 pandemic saw this methodology employed globally with mixed results \cite{Lewis2020Why}. Countries like China, South Korea, and Singapore have been lauded for their effective contact-tracing efforts \cite{Lancet2020Contact}, while countries such as the United Kingdom and the United States faced challenges in executing successful programs \cite{smith2020factors, Lancet2020Contact,Wang2022Effectiveness}. Despite the targeted nature of contact tracing, which avoids the broad societal and economic impacts of more blanket measures like school closures and travel bans, it is not without significant costs  \cite{Galea2020Mental}. Implementing these programs can be resource-intensive and may lead to unintended consequences, particularly regarding privacy when digital tracking systems are involved \cite{Lancet2020Contact}. Such concerns emphasize the need for a reasonable evaluation of the trade-offs associated with contact tracing initiatives.

The effectiveness of any public health intervention cannot be divorced from the societal context in which it is applied. To evaluate the success of both pharmaceutical and non-pharmaceutical interventions, we must take into account the network structure of social interactions and health behaviors within the population \cite{rizi2022epidemic, hiraoka2022herd, sajjadi2022structural, hiraoka2023strength}. Given the complexity of social structures, a strategy effective in one setting may fail in another. It is, therefore, imperative to rigorously evaluate the factors affecting the efficacy of contact tracing and other interventions, considering the diverse ways social structures can influence disease transmission.

The effectiveness of contact tracing is typically evaluated based on the number of infected individuals preemptively quarantined and its influence on halting transmission chains \cite{Kucharski2020Effectiveness,Wang2022Effectiveness}. This process is often assessed with the assumption that contact networks are tree-like. However, social networks consist of overlapping groups such as families and workplaces. Within these networks, an infected individual transmits the infection to specific group members, while contact tracing preemptively isolates others. Its success is most notable in the intersection of these groups—those who are both infected and isolated—as this effectively disrupts direct transmission chains. However, preemptively isolating uninfected members of these groups can also be crucial in controlling the spread of the disease. If contact tracing is not entirely effective, omitting some infectious individuals, the isolation of others becomes vital in stopping further infections. Accordingly, even isolations that might seem unnecessary due to contact tracing can significantly positively impact controlling the disease.

Social networks exhibit diverse and dense sub-structures which significantly impact contagion dynamics~\cite{feld1981focused,bokanyi2022anatomy}. These networks often feature clustering, crucial in complex contagion models for behavior spreading, where repeated exposure increases behavior adoption likelihood~\cite{keating2022multitype, Keating_generating}. This approach contrasts with traditional disease-spreading models that treat each infection event as independent. Gatherings can be modeled to show a nonlinear relationship between infected contacts and infection risk \cite{st2021universal}. Empirical studies have also highlighted that the effectiveness of contact tracing varies with the size of gatherings and can exhibit non-monotonic patterns \cite{mancastroppa2022sideward}. Theoretical works in this area include the development of a \textit{prompt} quarantine model in clique-based networks, where infected individuals and their contacts are quarantined with a fixed probability. As detailed in \cite{valdez2023epidemic}, this model results in continuous and discontinuous phase transitions and even backward bifurcations, offering new insights into epidemic control. It has also been shown that contact tracing is more effective for large-scale epidemics with low tracing rates in degree-assortative networks \cite{kiss2008effect}. In contrast, in disassortative networks, higher contact rates make it more effective. Disassortative networks are also more conducive to contact tracing for more minor epidemics due to the robustness of assortative networks against link removal \cite{kiss2008effect}. This underscores the complex interplay between network structures and epidemic control strategies, highlighting the need for tailored approaches in different network settings.

In this work, we demonstrate the group dynamics of contact tracing and their effect on outbreak sizes and the epidemic threshold by developing a stylized contact tracing model and a random network model with social groups manifested as cliques. This allows us to build on methods developed for spreading processes in networks with cliques \cite{salkind2008encyclopedia, luce1949method, stegehuis2021network}.  Our findings indicate that group structure enhances the effectiveness of contact tracing. Specifically, contact tracing in a network with cliques has a non-linear impact on the efficiency of halting the spread of chains that occur over a single link. This contrasts with models that assume a locally tree-like contact structure. We show that the combination of disease spreading and contact tracing can be approximated as a complex contagion process, where repeated exposures reduce the probability of infection because they can lead to isolation and thus can make subsequent infections impossible. This interpretation of contact tracing as a complex contagion explains our results on the importance of group structure.

The structure of this paper is organized as follows: In Section~\ref{sec:Model}, we introduce (A) the random network models employed and (B) detail the epidemic model, along with the contact tracing procedures. Section~\ref{sec:results} is divided into two main parts: 
(A) The first part focuses on identifying the epidemic threshold and observing the phase transition in epidemic size in networks with cliques. This is achieved using multi-type branching processes, which provide mean-field solutions for the reproduction number.
(B) The second part examines how the sub-critical epidemic size grows with disease parameters and the sizes of cliques. Section~\ref{sec:Complex_contagion} discusses how contact tracing in networks with cliques can be interpreted as a complex contagion process. Finally, in Section~\ref{sec:conc}, we highlight the implications of our findings for understanding and mitigating disease spread in social networks.

\section{Model}
\label{sec:Model}
We first introduce a random network model featuring cliques as social groups in Section \ref{sec:Networks}. After this, in Section \ref{sec:SIRQ}, we present the stylized dynamics we use for modeling disease spreading and contact tracing. 

\subsection{Random Networks with Cliques}
\label{sec:Networks}

 \begin{figure}
    \includegraphics[width=.9\linewidth]{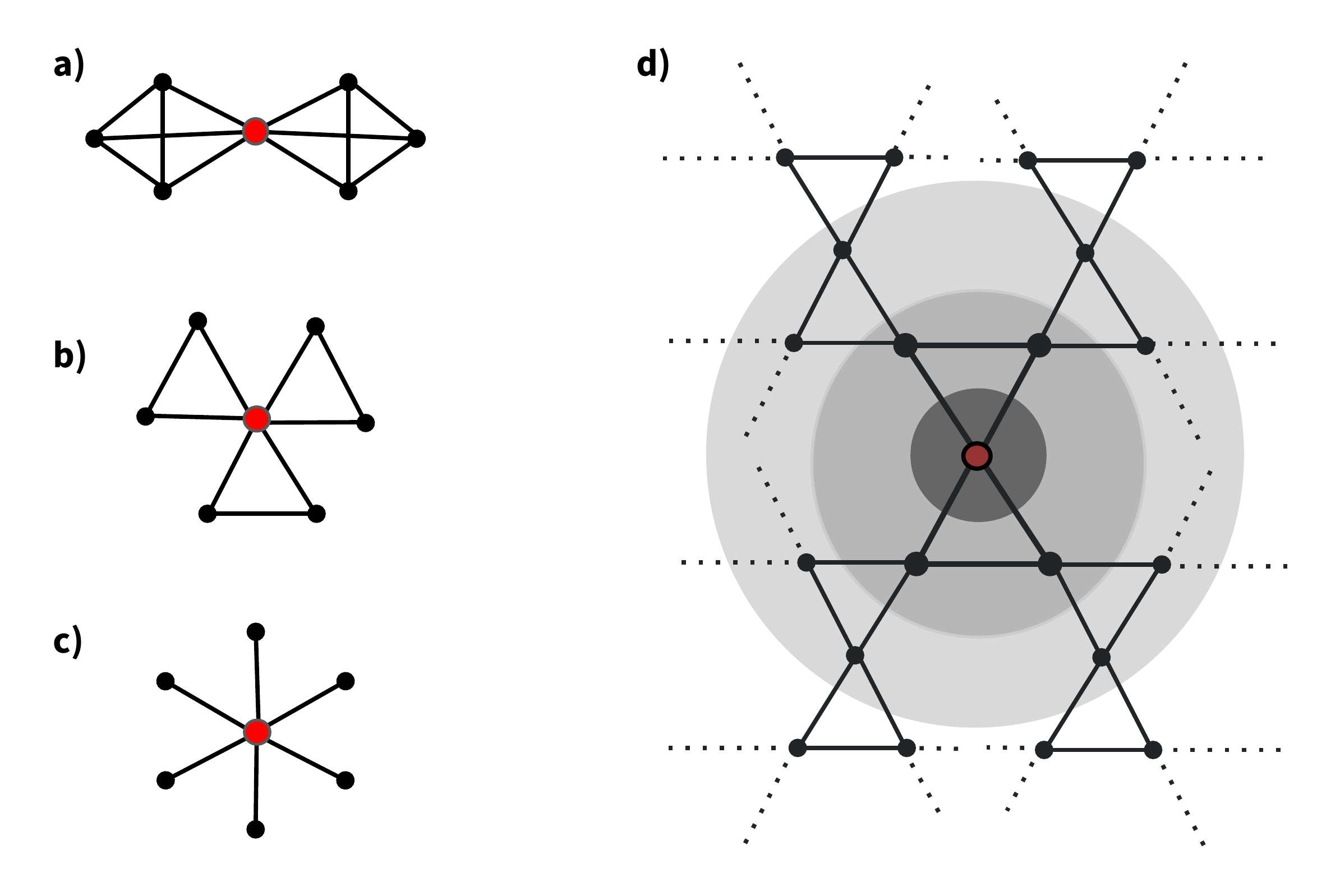}
    \caption{Illustration of $r$-regular $c$-clique network structures. Panels (a-c) highlight the immediate network vicinity of a focal (red) node within networks formed by 4, 3, and 2-cliques, respectively, where each node consistently has a degree of 6. These configurations are representative of the local topology repeated throughout the entire network. Panel (d) provides an example of a $4$-regular $3$-clique network, with each node having a degree of 4 and being part of two 3-cliques. Displayed are link stubs indicating connections to other nodes, demonstrating the typical local structure one would encounter in an extensive clique-based network. The shaded circular regions signify the proximity to a central node, which is marked in red. This shows the connectivity structure we examine using our $r$-regular clique-type networks. }
\label{fig:clique_networks}
\end{figure} 

Social networks, known for their complex, dense local structures, significantly differ from tree-like topologies, especially in disease-spreading scenarios \cite{osullivan_mathematical_2015}. This complexity is due to high clustering in social units like families and workplaces \cite{luce1949method, feld1981focused, karrer2010random, bokanyi2022anatomy}, leading to the need for novel tools to understand cliques' effects on spreading processes \cite{karrer2010random, keating2022multitype}.
We aim to investigate the sole effect of social groups on contact tracing, ignoring other salient social network features such as degree heterogeneity or homophily \cite{rizi2022epidemic, hiraoka2022herd}. 

In studying epidemic processes on networks, cliques are idealized representations of social groups within contact networks. Each social group is represented as a complete graph, where every member is connected to every other member, illustrating the all-to-all connection pattern within these groups. Panel (a) and (b) in Fig.~\ref{fig:clique_networks} depict a focal node that belongs to two 4-cliques and three 3-cliques, respectively. In social networks, $c$-cliques are complete subgraphs representing a group of $c$ people who are all connected and, thus, can potentially infect each other \cite{salkind2008encyclopedia, luce1949method}. 

To investigate the sole impact of group structure, we compare network results in which nodes possess an equal number of connections but belong to groups of different sizes. In our model, we analyze homogeneous networks where all nodes have the same number of links and the same level of network clustering. We then compare different homogeneous networks, where only the amount of clustering varies between networks. This will allow us to isolate the impact of contact tracing on spreading disease in the presence of cliques. In practice, this is made possible using clique-based network generation methods \cite{osullivan_mathematical_2015, miller2009percolation, PhysRevE.68.026121, kiss2008comment, PhysRevE.82.036115, keating2022multitype}.

The algorithm for generating networks with prescribed clique structures generates an auxiliary bipartite graph with one part for individuals and the other for groups, where links are only between individuals and groups. This network is then projected into a unipartite network of individuals, where the groups become cliques, i.e., all the individuals in each group are connected.
We construct the bipartite graph by giving each group $c$ stubs and each individual $n_c$ stubs. As we have a total of $N$ individuals, each with degree $n_c$, the total number of stubs leading out of individuals should be $N \times n_c$. The total number of stubs leading out of groups must be the same. Therefore, the bipartite network must have $(N \times n_c)/c$ groups. We connect, uniformly at random, stubs leading out of individuals to stubs leading out of groups. This provides a bipartite network that defines which individuals belong to which groups. Finally, we take an unipartite projection of the bipartite network where individuals are the only nodes we keep, and we connect two individuals if they belong to the same group, i.e., if they are connected to the same group in the bipartite network. The groups become the cliques in the contact network, connecting individuals who are the nodes.

In the thermodynamic limit, such contact networks have a vanishingly small number of self-loops or multi-links. See \cite{wormald1999models, bollobas1998random} for further details on these network structures. In practice, when we build such networks for our simulations, we remove the few self-loops and multi-links, ending up with a simple graph. Note that each $c$-clique contributes $c-1$ links to a node degree. Therefore, the number of cliques that a node is a part of, $n_{c}$, satisfies the condition $n_{c}(c-1) = r$. When $c=2$, the model generates a random $r$-regular graph, see Fig.~\ref{fig:clique_networks}c. Fig.~\ref{fig:clique_networks}d illustrates a $4$-regular $3$-clique network, where each node has a degree of 4 and is part of two 3-cliques.

\subsection{SIRQ Dynamics}
\label{sec:SIRQ}
Both contact tracing and disease transmission are complicated processes in reality and are affected by various details related to the particular disease, contact tracing procedure, and the underlying social system. We aim to reduce these complications into a minimal mathematically tractable model that captures stylized dynamics of contact tracing and disease spreading. We employ a discrete-time Susceptible-Infectious-Recovered (SIR)  model to model disease dynamics \cite{pastor2015epidemic}, where at each time step, each infected ($\mathrm{I}$) individual independently infects each neighboring susceptible  ($\mathrm{S}$) node with \textit{transmission} probability $p$. After this, the infected individuals are moved to the recovered ($\mathrm{R}$) compartment. Importantly, this time-discretized model ignores variations in recovery times and can only implicitly consider complications such as incubation periods \cite{newman2002spread}. 

\begin{figure}
    \centering
    \includegraphics[width=.6\linewidth]{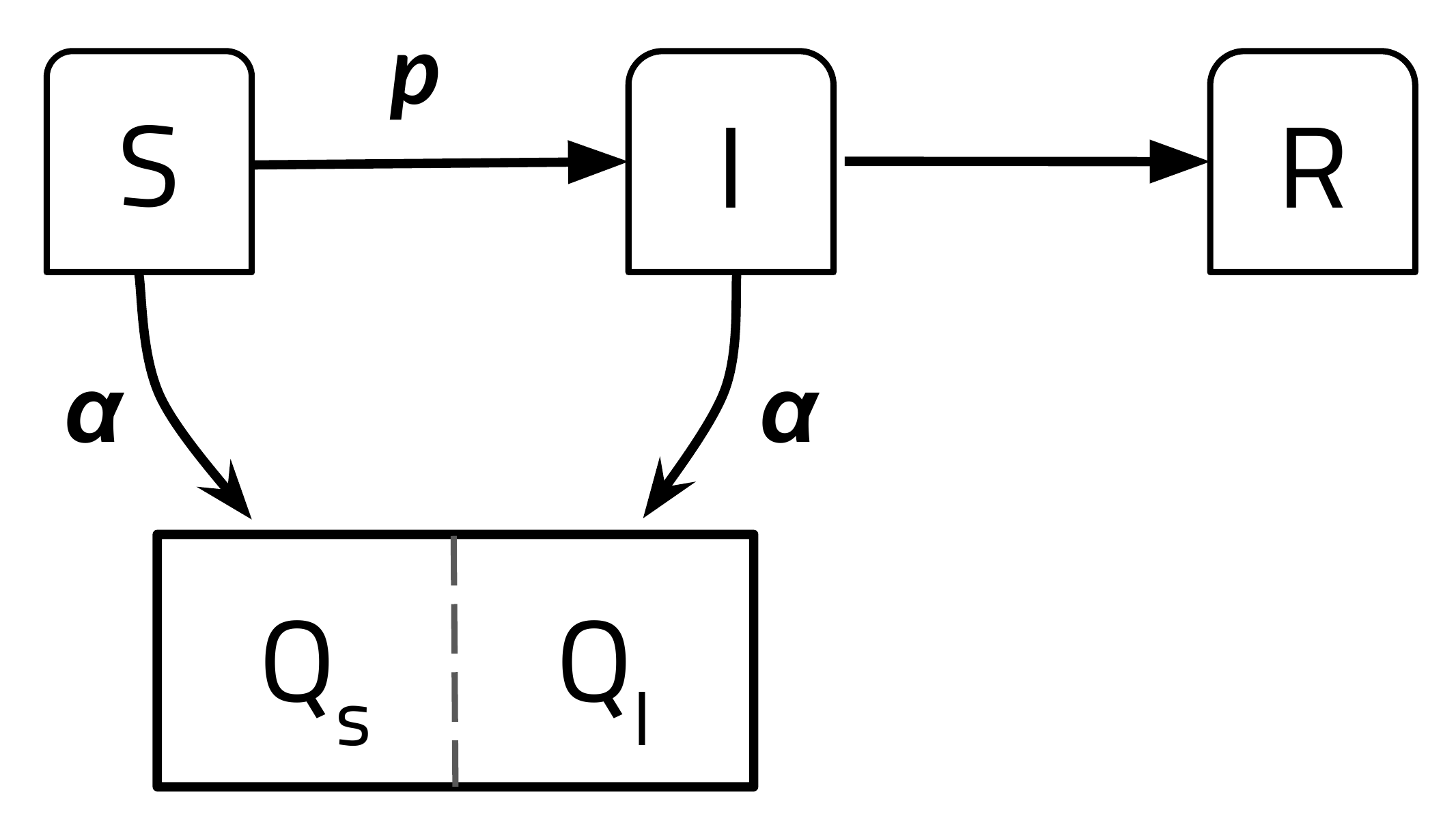}
    \caption{Diagram of the SIRQ model showing the flow between compartments based on transition probabilities based on the stochastic dynamics introduced in Sec.~\ref{sec:SIRQ}. Susceptible individuals become infected with probability $p$ and enter quarantine with probability $ \alpha $. The $\mathrm{Q}$ compartment includes people in quarantine, either infected or susceptible. Those who are both infected and quarantined move to the $\mathrm{Q_I}$  sub-compartment, while those who are only quarantined go to the $\mathrm{Q_S}$ sub-compartment of $\mathrm{Q}$. Fig.~\ref{fig:demo} depicts these two situations. Infected individuals who are not quarantined go to the $\mathrm{I}$ compartment and will recover deterministically in the subsequent time step.}
\label{fig:demo0}
\end{figure} 

\begin{figure}
    \includegraphics[width=.7\linewidth]{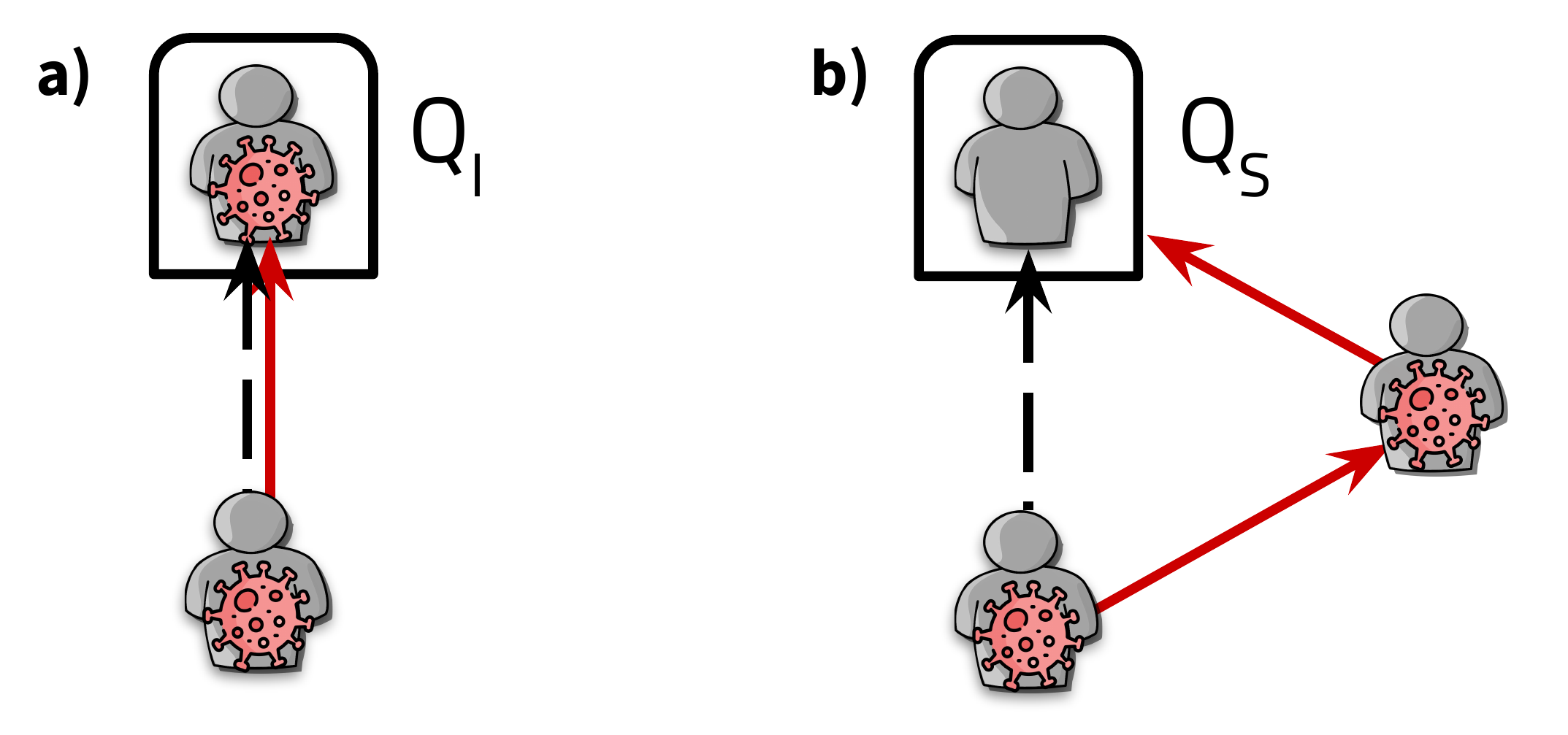}
    \caption{Schematic of contact tracing and spreading without loops (a) and with local loops (b). Infections that would be successful are marked with solid red links, and successful contact tracing with dashed black links. After each exposure, a susceptible node isolates itself with probability $\alpha$ and becomes infected with probability $p$ independently.
    If no loops are considered, the combination of infections and contact tracing can be reduced to a single link. There are four possible scenarios: nothing happens; the infection spreads to the neighbor, but contact tracing fails; the infection fails to spread, but contact tracing succeeds; or (a) both infections spread, and the contact tracing succeeds so that the node will be in sub-compartment $\mathrm{Q_I}$. The last case is where we can benefit from contact tracing cutting indirect spreading paths thanks to the presence of clustering. 
    (b) With local loops, an infection through a common neighbor of both nodes can be avoided. As the quarantine takes place close to the infection, it can prevent the infection from arriving at the neighbor through a local loop as the node is in sub-compartment $\mathrm{Q_S}$.}
\label{fig:demo}
\end{figure}

Contact tracing can be implemented in various ways, such as with phone applications \cite{rizi2022epidemic}, in different manual tracing settings, or with combinations of these two \cite{zhang2022epidemic}. The success of contact tracing can be affected by the ability of individuals to recall contacts, the delay times in the tracing process, mobile phone application adoption, and the extent to which the individuals follow the isolation or quarantine recommendations \cite{anglemyer8bero, fetzer2021measuring, hossain2022effectiveness}. We model all these complications with the probability $\alpha$ of a neighboring node successfully moving to compartment $\mathrm{Q}$ such that all further infections are avoided. Further, the contact tracing moving nodes to the $\mathrm{Q}$ compartment is done independently using the same contact network as the infections. In the model, this translates to each infected node placing each neighboring node into compartment $\mathrm{Q}$ with probability $\alpha$. 
The nodes in the $\mathrm{Q}$ compartment can be either infected ($\mathrm{Q_I}$) or susceptible ($\mathrm{Q_S}$). It is important to highlight that the nodes within $\mathrm{Q}$ are distinctly separated from those in set $\mathrm{R}$, as they are housed in separate compartments. Despite this distinction, it should be noted that neither group of nodes contributes to the propagation dynamics. Fig.~\ref{fig:demo0} depicts the compartmental structure and the associated transition probabilities, while Fig.~\ref{fig:demo} demonstrates the benefits of contact tracing, particularly when loops are present. 

Our model treats isolation and quarantine identically, encapsulating both by the probability $\alpha$. Consequently, we will refer to both terms interchangeably, reflecting their similar dynamics in our model. In public health, however, isolation and quarantine are distinct strategies for preventing the spread of contagious diseases \cite{us2020difference}. Isolation separates individuals who are sick with an infectious disease from those who are not, and it is applied to confirmed cases to prevent the spread of the infection to others. Quarantine, in contrast, involves separating and restricting the movement of people exposed to a contagious disease to see if they become sick, targeting those who may have been exposed but are not yet confirmed to be ill. Thus, in public health literature, while isolation is for those already sick and contagious, quarantine is for those who might become sick due to exposure \cite{us2020difference}. It is worth mentioning that our model also captures the impact of Ring Vaccination \cite{kucharski2016effectiveness}, a strategy that involves vaccinating individuals around an infected person, effectively isolating them from the disease network \cite{kretzschmar2004ring}. This strategy, which successfully eradicated smallpox \cite{schleiff2020multi}, is paralleled in our model by transitioning individuals to a quarantined compartment with probability $\alpha$.

Our model assumes the infection and contact tracing processes are independent (i.e., we treat $p$ and $\alpha$ as independent probabilities). See Fig.~\ref{fig:demo} for an illustration of the process. The order in which they are evaluated in the discrete-time model does not make a difference for the epidemic threshold. However, there is a slight variation in the epidemic size depending on the order, as the number of isolated infected individuals is affected by the order in which nodes are infected and placed in the quarantined compartment. 
For this purpose, we follow an order where we go through one infected-susceptible link at a time. First, we evaluate the spread of the epidemic and then the contact tracing for that link. 

We focus on contact tracing when the disease does not reach a significant part of the population. An upsurge in the number of infections can strain the contact tracing process, leading to increased delay times that potentially weaken its overall effectiveness \cite{ferretti2020quantifying, zhang2022epidemic}. In reality, a node may be reinfected after leaving the compartment Q. However, as our contact networks are large enough random graphs with cliques (see Sec.~\ref{sec:Networks}), the infection paths will not form significant long loops. This means that for our model, we can assume that the isolation times are long enough that they will stop all the incoming infections to a node. We can indefinitely keep the isolated nodes in the $\mathrm{Q}$ compartment. In other words,
with prolonged isolation, which may vary based on clique size, post-quarantine infection becomes negligible. 
Intuitively, this modeling choice can be understood as the re-entry of quarantined individuals into the susceptible ($\mathrm{S}$) or infected ($\mathrm{I}$) states being unlikely before the infection subsides locally. The impact of this assumption is explored in more detail in Appendix~\ref{sec:shortQtime}.

\section{Results}
\label{sec:results}
In this section, we demonstrate the impact of contact tracing in networks with cliques. We begin by analyzing epidemic thresholds in tree-like networks (Sec.~\ref{sec:tree_R}) and then assess phase transitions in epidemic sizes in networks with cliques (Sec.~\ref{sec:SIR}). We employ a multi-type branching process to understand the influence of spreading parameters and clique sizes on the effective reproduction number (Sec.~\ref{sec:m-field}). The effect of contact tracing on these thresholds is examined (Sec.~\ref{sec:m-field_epi_thr}), followed by an analysis of outbreak sizes in sub-critical regimes, highlighting the role of quarantine probability and clique size (Sec.~\ref{sec:Outbreak_Size_Revisited}).

\subsection{Epidemic Threshold \& Reproduction Number}
\label{sec:R0}
Given a population in a demographic steady state, with no history of a given infection or introduction of any intervention, the\textit{ basic reproduction number} $R_0$ determines if the introduction of the infectious agent causes an outbreak ($R_0 >1$) or not ($R_0 < 1$) in the absence of interventions \cite{diekmann2013mathematical}. This is because $R_0$ yields the expected number of secondary cases produced by a typical infectious individual throughout their contagious period in a fully susceptible population. Therefore, $R_0$ as a threshold for the stability of a disease-free equilibrium in a compartmental model divides the phase space into super/sub-critical regions, respectively. When interventions such as contact tracing are implemented, we use the term \textit{effective} reproduction number $R_\mathrm{e}$ instead of the \textit{basic} reproduction number to differentiate between situations with no interventions in this paper. Therefore, to determine if the epidemic dies out or yields an outbreak in the presence of an intervention, we need to compute the value of $R_\mathrm{e}$ \cite{hiraoka2022herd}. $R_\mathrm{e}$ as a bifurcation parameter in our epidemic model depends on the spreading parameters, $p$ and $\alpha$, and the network structure, which is determined by $c$ and $r$.  

\subsubsection{Random Tree-like Networks}
\label{sec:tree_R}
For a large tree-like network, like a random $r$-regular graph built with blocks the same as the one in Fig.~\ref{fig:clique_networks}c, we can find the epidemic threshold in the $\alpha p$--plane using
\begin{equation}
    R_\mathrm{e} = p(1-\alpha)\bar{d},
\label{eq:critical}
\end{equation}
and setting $R_\mathrm{e} = 1 $. Here, $\bar{d}$ represents the average excess degree of the network. This is the average count of additional connections that a node has, apart from the one used to arrive at when it is found by traversing a uniformly randomly selected link in the network. As our networks have uniform degree distributions, such that every node has degree $r$, the expected excess degree is just the degree minus one, $\bar{d} = r - 1$. It should be noted that even with the most severe disease with $p=1$, it is still possible to avoid an outbreak. By setting $p=1$ in Eq.~\ref{eq:critical} and solving for $\alpha$, we can determine that if the quarantine is carried out in such a way that $\alpha>1-1/\bar{d}$, the effective reproduction number, $R_\mathrm{e}$, will remain below 1. 

In general, for a tree-like random network with expected excess degree $\bar{d}$, we can rewrite the effective reproduction number as a product of spreading properties and network structure as $R_\mathrm{e}=p_{\mathrm{e}}\bar{d}$ where $p_{\mathrm{e}}$ is the \textit{effective transmission probability} and defined as
\begin{equation}
    p_{\mathrm{e}} = p(1-\alpha).
    \label{eq:effective transmission probability}
\end{equation}
So every active node can, on average, infect $R_\mathrm{e} = p_{\mathrm{e}}\bar{d}$ new people who can propagate the disease, i.e., are not themselves quarantining. When contact tracing is not in place ($\alpha = 0$), the effective reproduction number reduces to the basic reproduction number, $R_0 = p\bar{d}$. Since we ignore variations in recovery time, the SIR dynamics can be mapped to a bond percolation problem, where $p$ represents the link occupation probability and the size of the giant component corresponds to the final outbreak size \cite{newman2002spread, kenah2007second, kenah2011epidemic, hiraoka2022herd}. This mapping results in the epidemic threshold being equivalent to the percolation threshold, which occurs at $p_{*} = {1}/{\bar{d}} = {1}/{(r-1)}$. Thus, a phase transition is expected from a disease-free equilibrium to an endemic state.
 For example, for a 6-regular graph $p_{*} = 0.2$, and
when $\alpha = 0.5$, the epidemic threshold occurs at $p = 0.4$, according to Eq.~\ref{eq:critical}. Our forthcoming explanation of simulation results of random networks with cliques shows that this equation aligns exceedingly well with $c=2$ (no loops). When the tree-like assumption does not hold, for example, when $c>2$ in our network model, an alternative method is required to determine $R_\mathrm{e}$. This is the focus of the following sections.

\subsubsection{Simulating the Epidemic}
\label{sec:SIR}
Eq.~\ref{eq:critical} is not applicable for networks with cliques, as it assumes a locally tree-like structure. However, we can simulate the epidemic dynamics to observe how the outbreak size varies with changes in disease parameters and network structure. These simulations reveal a sharp increase in the outbreak size, transitioning from a few individuals to a significant portion of the network, upon crossing certain thresholds of $p$ and $\alpha$. Additionally, at the epidemic threshold, an ensemble of simulations shows considerable variation in outbreak sizes, reflecting the critical nature of this point as noted in \cite{li2021percolation, kenah2011epidemic}.
In our simulations, we build large networks (with $N \approx  10^5$ nodes) according to Sec.~\ref{sec:Networks}, run the SIRQ dynamics $10^4$ times, and find the ensemble average of the number of nodes in the different compartments in each run as our measure of disease spread from simulations. Using this, we can calculate other quantities of interest. For example, the size of the outbreak is then given by the ensemble averages of the number of people in the $\mathrm{R}$ and $\mathrm{Q_I}$ compartments. 
We follow an order to go through one infected-susceptible link at a time for the epidemic size computations. First, we evaluate the spread of the epidemic and then the contact tracing for that link. This way, the number of infected people in quarantine can be computed as $N_{\mathrm{Q_I}}=pN_{\mathrm{Q}}$. We find the size of an epidemic $E$ by summing up the number of infected people in and out of quarantine, $N_{\mathrm{Q_I}}$ and $N_{\mathrm{R}}$, respectively. 

 \begin{figure}
\includegraphics[width=.5225\linewidth]{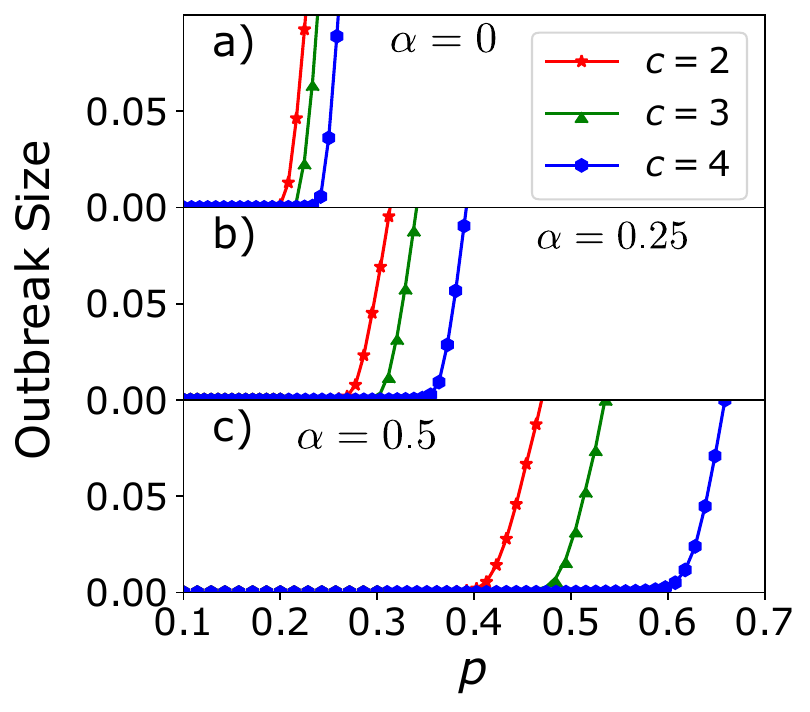}%
\includegraphics[width=.4775\linewidth]{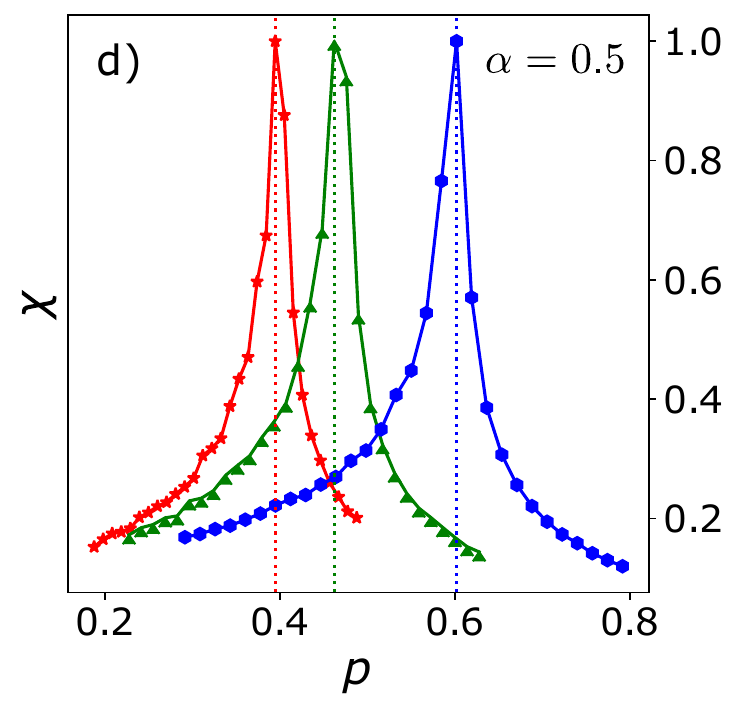}
    \caption{Phase transitions from a disease-free equilibrium to an endemic state for 2, 3, and 4-clique networks with degree 6 as introduced in Sec.~\ref{sec:Networks}. (a-c) The outbreak size $E$, normalized to the network size, is shown on the vertical axis for when (a) $\alpha = 0$ (no contact tracing), (b) $\alpha = 0.25$, and (c) $\alpha = 0.5$, from top to bottom respectively. Note that the transition points are shifted slightly to the right for larger clique sizes, $c$, even when there is no contact tracing ($\alpha=0$), but this difference is substantially amplified for larger $\alpha$ values. (d) The coefficient of variation of outbreak sizes in an ensemble, $\chi$ normalized to unity for $\alpha=0.5$. We use $\chi$ to numerically detect the transition point as it peaks at the epidemic threshold. Results are based on Monte Carlo simulations introduced in Sec.~\ref{sec:SIR}.}
    \label{fig:phase-transition}
\end{figure}

Fig.~\ref{fig:phase-transition}(a-c) illustrates the dependence of outbreak size on the value of $p$ under different scenarios, namely in the absence of contact tracing ($\alpha = 0$), and with contact tracing at $\alpha = 0.25$ and $\alpha = 0.5$, for networks consisting of 2, 3, and 4-cliques. As the clique size increases, the outbreak size decreases for any given transmission probability. Moreover, this effect is magnified by an increase in the value of $\alpha$.  

Furthermore, we use the fluctuations in the outbreak sizes, $\chi$, for determining the epidemic thresholds as illustrated in Fig.~\ref{fig:phase-transition}d. Fluctuation in outbreak sizes typically displays a peak even in finite systems. When computed as a function of infection probability $p$, the peaks in $\chi$ indicate the epidemic thresholds for some value of $\alpha$. This measure is analogous to susceptibility in Critical Phenomena \cite{goldenfeld2018lectures}, which measures the response magnitude generated by a small external field disturbance \cite{dorogovtsev2008critical}. In practice, we run a set of simulations and calculate the coefficient of variation of the outbreak sizes, which is the ratio of the standard deviation of outbreak sizes to their ensemble average, $\chi = {\sigma_E}/{\langle E \rangle}$ \cite{pastor2015epidemic}.
 Fig.~\ref{fig:phase-transition}d shows that for a fixed $r$, here $r=6$, contact tracing on networks with cliques is more effective when the contact networks include larger cliques. So, the larger the clique size, the larger the critical transmission probability.

\subsubsection{Mean-filed Reproduction Number}
\label{sec:m-field}

\begin{figure}
    \includegraphics[width=\linewidth]{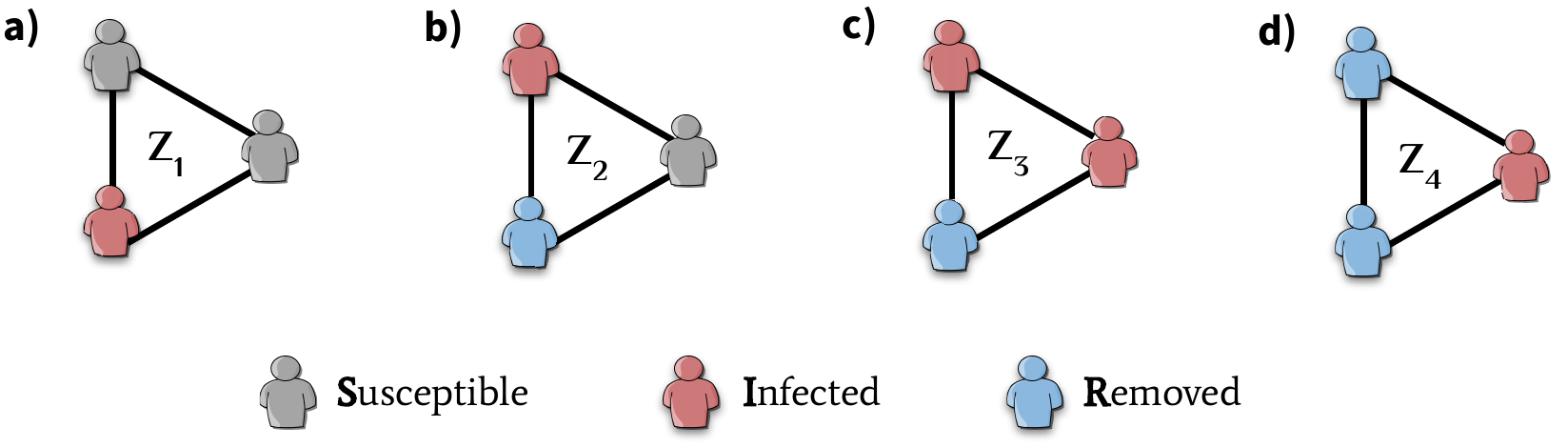}
    \caption{Each 3-clique can have four life stages or diffusion patterns with at least one infected node. Section~\ref{sec:m-field} considers both recovered and quarantined nodes in the $\mathrm{R}$ compartment. Using a 6-regular 3-clique network, we observe that a $Z_1$ node can form a $Z_2$ motif with two $Z_1$ nodes and a $Z_3$ motif with four $Z_1$ nodes. A $Z_2$ node can also create a $Z_4$ motif with two $Z_1$ nodes. Nodes in the $Z_1$, $Z_2$, and $Z_4$ motifs can transition to an infection-annihilated states such as $\{\mathrm{R}, \mathrm{S}, \mathrm{S}\}$ which are not shown here. In Section~\ref{sec:Complex_contagion}, we assume that both quarantined and susceptible nodes are in the $\mathrm{S}$ compartment, while only recovered individuals are in the $\mathrm{R}$ compartment.}
\label{fig:clique_states}
\end{figure} 

By representing our stylized SIRQ model using a multi-type branching process, we can derive the relationship between $p$, $\alpha$, and clique size, $c$, on the epidemic threshold. In the multi-type branching process representation of our SIRQ model, we track different clique states, which we refer to as \emph{clique motifs}. Each clique motif accounts for the possible number of susceptible, infected, or recovered nodes that any clique can inhabit at a given time. Every possible motif is denoted by $Z_i$ (refer to Fig.~\ref{fig:clique_states} for listing possible clique motifs for a 3-node clique). Regardless of the network structure, we can always average the expected number of new infections over all possible infected types from our multi-type branching process with the next-generation matrix \cite{diekmann2013mathematical}. To do this, we track the propagation of clique motifs in a network under the introduced dynamics and form the next-generation matrix $\mathbf{M}$ for the number of motifs in the network. The matrix $\mathbf{M}$ is also known as the mean matrix or the population projection matrix \cite{caswell2000matrix}, and its element $m_{ij}$ gives the expected number of motifs of type $Z_i$ that are created in the next time steps from a motif of type $Z_j$. 

Fig.~\ref{fig:clique_states} shows the motifs corresponding to the four life stages of a 3-clique. In this representation, we have combined the isolated and recovered nodes into a single R compartment because these two compartments are equivalent for the epidemic threshold computations. The next-generation matrix represents the transitions between these motifs. For example, the infected node in $Z_1$ can infect one or two neighbors, corresponding to motifs $Z_2$ and $Z_3$. Further, $n_c-1$ new $Z_1$ motifs are produced every time such an infection occurs. That is, when $Z_1$ turns into $Z_2$, there are also $n_c-1$ new $Z_1$ motifs, and when it turns into $Z_2$, there are $2(n_c-1)$ new $Z_1$ motifs created.

Table~\ref{table:mper} shows the non-zero elements of $\mathbf{M}$. For example, the transition from $Z_2$ to $Z_4$ occurs when contact tracing fails (with probability $1-\alpha$). The infection is successful with probability $p$, which means that, in expectation, a single $Z_2$ motif produces $m_{42} = p(1-\alpha)$ new $Z_4$ motifs. The motif $Z_4$ can also be made when the infected node in $Z_1$ puts one neighbor in quarantine (with probability $\alpha$) and fails to do so for the neighbor and infects it instead (which happens with probability $p(1-\alpha)$). As there are two ways of choosing the infected and isolated neighbor, the expected number of $Z_4$ motifs produced by the $Z_1$ motif is given by $m_{41} = 2\alpha[(1-\alpha)p]$.
The rest of the transitions are produced similarly by computing the probabilities of going from one motif to another. As described in Appendix~\ref{sec:auto_matrices}, we write general formulas for any transition and use this to automatically generate the desired mean matrix, $\mathbf{M}$, for cliques of any size. 

\begin{table}[h]
\begin{tabular}{ c c c } 
 \hline
 \hline
$i,j$ &\hspace{3mm} &$m_{ij}$ \\
 \hline
$ 1 , 1 $&&$ 4 p \left(1- \alpha \right) $\\
$ 1 , 2 $&&$  2p \left(1-\alpha \right) $\\
$ 2 , 1 $&&$ 2 p \left(1-\alpha \right)^{2} \left( 1-p\right) $\\
$ 3 , 1 $&&$ p^{2} \left(1-\alpha\right)^{2} $\\
$ 4 , 1 $&&$ 2 \alpha p \left(1 - \alpha\right) $\\
$ 4 , 2 $&&$ p \left(1 - \alpha\right) $\\
 \hline
 \hline
\end{tabular}
\caption{Non-zero elements of the next-generation matrix $\mathbf{M}_{4\times 4}$ for a 3-clique network. $m_{ij}$ gives the expected number of $Z_i$ cliques from a $Z_j$ clique, as shown in Fig.~\ref{fig:clique_states}.}
\label{table:mper}
\end{table}

\begin{figure}
\includegraphics[width=\linewidth]{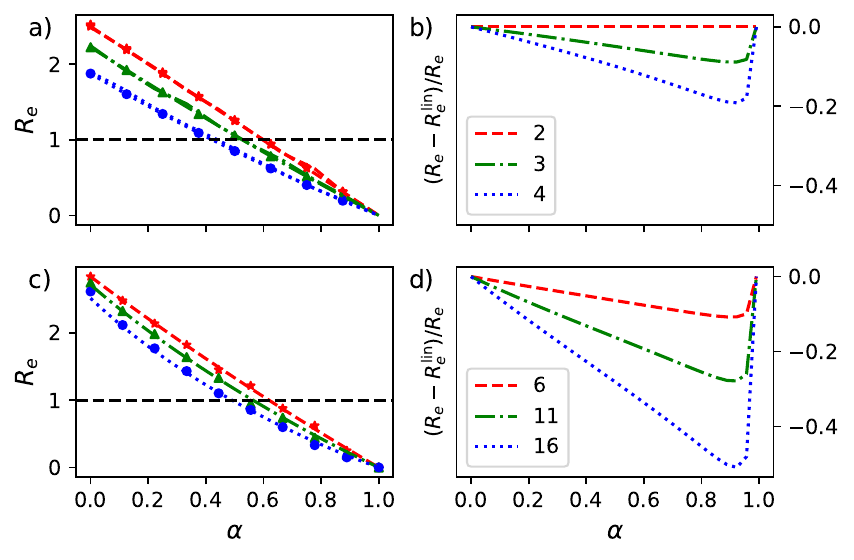}
\caption{The impact of contact tracing in clique networks on mitigating epidemic spread: This figure illustrates the decline in the effective reproduction number, $R_\mathrm{e}$, with the contact tracing parameter, $\alpha$, across networks with various clique configurations. Specifically, we examine cliques of sizes $c = 2, 3, 4$ with $r = 6$ and transmission probability $p = 0.5$ in Panel (a), and cliques of sizes $c = 6, 11, 16$ when $r = 30$ and $p = 0.1$ in Panel (c). Networks with larger cliques achieve the critical threshold of $R_\mathrm{e} = 1$ with less contact tracing effort. 
 When $c=2$, the influence of contact tracing on $R_\mathrm{e}$ aligns linearly with $\alpha$, according to Eq.~\ref{eq:critical}. In scenarios involving larger cliques, this relationship turns concave and is further intensified as the clique size increases or the transmission probability decreases. The dotted lines are from the mean-field calculations introduced in Sec.~\ref{sec:m-field}, and the markers are from Monte Carlo simulations described in Sec.~\ref{sec:SIR}. Panels (b) and (d) show a relative difference of $R_\mathrm{e}$ to the linear case when $c=2$ ($R^\mathrm{lin}_\mathrm{e}$). Fig.~\ref{fig:r0_old} shows similar results to panel (a) for different $p$ values. The larger the transmission probability, the larger the differences between the curves of other networks with cliques.}
\label{fig:r0}
\end{figure}

What is significant about the next-generation matrix is that its spectral radius (Perron root), or the largest modulus of the eigenvalues \cite{varga1999matrix}, yields the effective reproduction number \cite{Diekmann1990Definition, brouwer2022spectral} such that
\begin{equation}
    R_\mathrm{e} = \rho(\mathbf{M}),
    \label{eq:specteral}
\end{equation}
and epidemic thresholds for any given clique network can be found for finding $p$ and $\alpha$ such that $R_\mathrm{e} = 1$. We give more detailed arguments about this identity in Appendix~\ref{sec:irreducibility}, and show that this definition aligns with the simulation results of Fig.~\ref{fig:r0}.

In Fig.~\ref{fig:r0}, we present the effective reproduction number, $R_\mathrm{e}$, across various clique sizes under differing transmission probabilities, integrating results from both mean-field calculations and simulations, detailed in Sec.~\ref{sec:r0_old}. The figure reveals a non-linear relationship between $R_\mathrm{e}$ and the intervention parameter $\alpha$, particularly for larger cliques. This observation is crucial in practical scenarios where the basic or effective reproduction number is a key metric for monitoring and controlling epidemic situations. Public health officials often rely on this data, represented on the horizontal axis, to gauge the extent of interventions required to bring the epidemic under control, aiming to reduce $R_\mathrm{e}$ below the critical threshold of 1 by increasing $\alpha$. 

More successful contact tracing leads to a lower effective reproduction number, and the extent of this reduction in networks with cliques is larger than in tree-like networks. As seen from Eq.~\ref{eq:critical}, given the basic reproduction number, $R_0$ ($R_\mathrm{e}$ without contact tracing), the $R_\mathrm{e}$ decreases linearly with $\alpha$, following $R_\mathrm{e}^{\mathrm{lin}} = (1- \alpha ) R_0$. However, as demonstrated in Fig.~\ref{fig:r0}(a,c), this reduction is not linear in networks with cliques. The difference between this simplistic linear estimation and the more realistic $R_\mathrm{e}$ (Eq.~\ref{eq:specteral}) yields the error of assuming contact networks are locally tree-like. Fig.~\ref{fig:r0}(b,d) higlight this discrepancy by showing the relative error, $\frac{R_\mathrm{e} - R_\mathrm{e}^{\mathrm{lin}}}{R_\mathrm{e}}$, between these two approaches. These errors become more pronounced in networks with larger groups. It reaches around $20\%$ for cliques of size 4 and around $50\%$ for cliques of size 16. 

\subsubsection{Epidemic Threshold}
\label{sec:m-field_epi_thr}
\begin{figure}
\includegraphics[width=.5\linewidth]{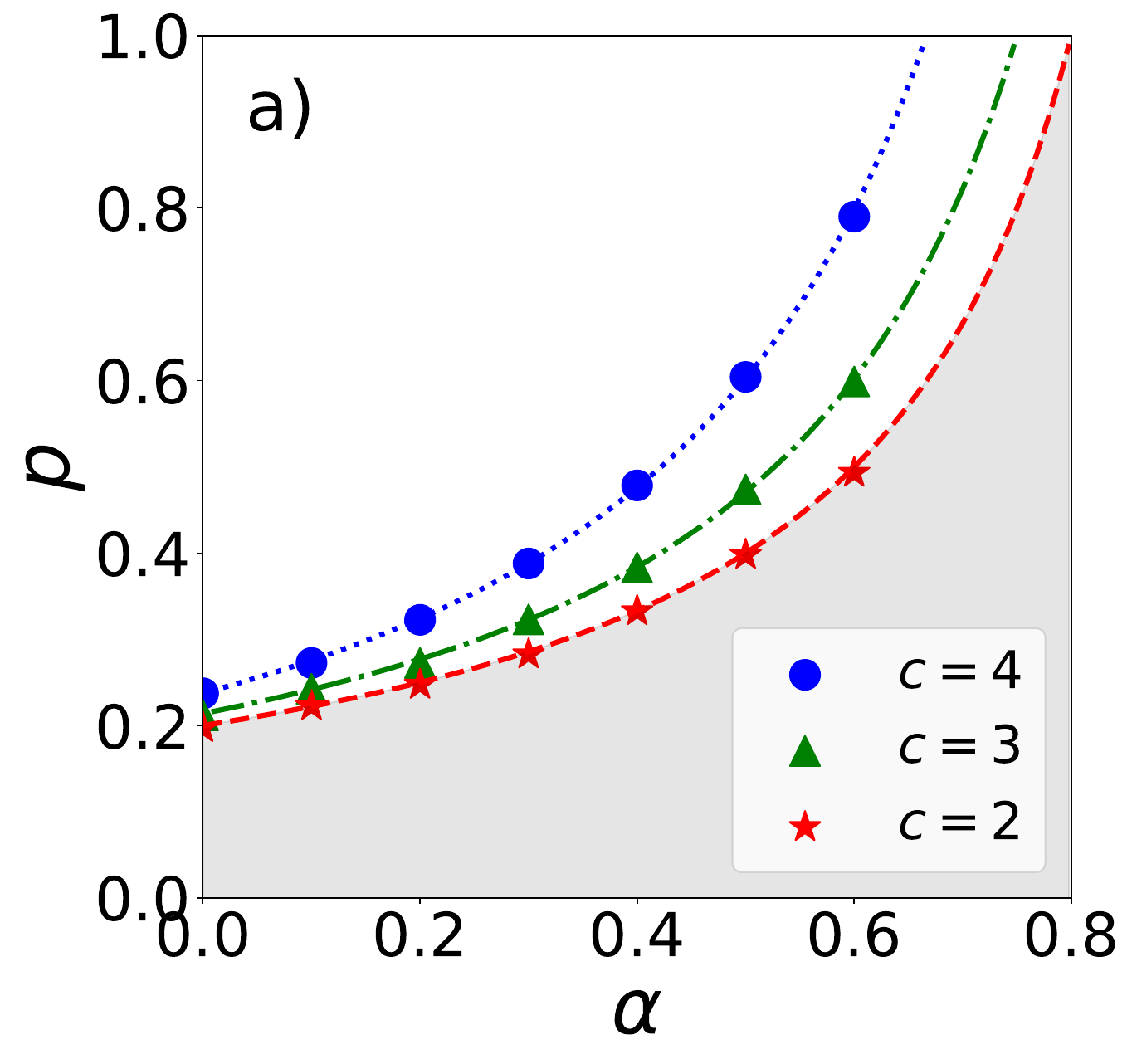}%
\includegraphics[width=.5\linewidth]{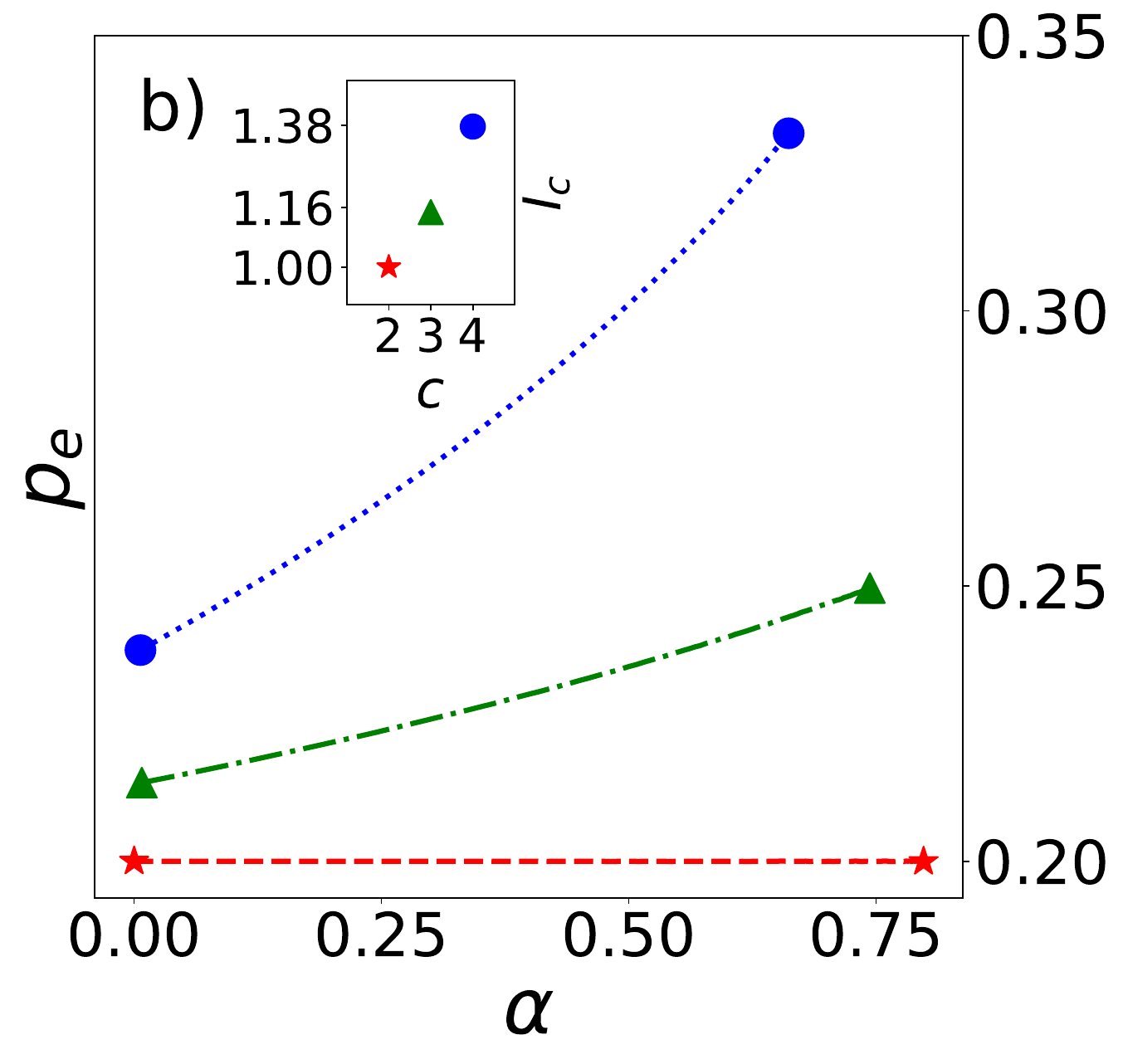}
    \caption{Phase diagram showing that increasing the clique size increases the epidemic threshold and effectiveness of contact tracing. 
    (a) The critical curves  where $R_\mathrm{e}=1$ in the $\alpha p$-plane for $c \in \{2,3,4\}$. The lines indicate the results of the mean-field approximations described in Sec.~\ref{sec:R0}, and the markers show results for simulations described in Sec.~\ref{sec:SIR}.
    The shaded area is the sub-critical region for $c=2$ where the infection eventually dies out after a finite number of generations for any clique size. 
    (b) The same phase diagram in $\alpha p_\mathrm{e}$-plane, where $p_\mathrm{e}=p(1-\alpha)$ is the effective transmission probability defined in Eq.~\ref{eq:effective transmission probability}. 
    The larger markers in the right end in panel b indicate extreme points from Eq.~\ref{eq:extremeP=1}.
    The inset in panel b shows the relative maximum increase in the effective epidemic threshold for different networks with cliques. Each point in the inset is the ratio of the $p_\mathrm{e}$ values at the endpoints of each curve outside the inset, such that $I_c = p_\mathrm{e}(\alpha_\mathrm{max})/p_\mathrm{e}(\alpha_\mathrm{min}) $.} 
\label{fig:critical_curve_theory}
\end{figure}

Next, We will use the mean-field framework developed above to investigate the epidemic thresholds. Fig.~\ref{fig:critical_curve_theory}a presents the phase diagram of the epidemic for various networks with cliques by drawing $R_\mathrm{e} = 1$ curves in the $\alpha p$--plane. These curves divide the plane into sub- and super-critical regions. In the sub-critical region, there is no possibility of an outbreak that scales with the network size. In contrast, in the super-critical region, there is a positive probability of such an outbreak upon infection. For networks with cliques, increasing the clique size enlarges the sub-critical region and shrinks the super-critical region. Note that even without contact tracing (i.e., when  $\alpha = 0$), including a clique structure in random graph models raises the epidemic threshold slightly. 
However, this effect is amplified by contact tracing, which can be observed as an increased difference between the critical $p$ value as $\alpha$ increases.

We illustrate that the combined impact of contact tracing and cliques is larger than one would expect by the tree-like assumption by plotting the critical effective transmission probability $p_\mathrm{e}$, defined in Eq.~\ref{eq:effective transmission probability}, as a function of $\alpha$. Fig.~\ref{fig:critical_curve_theory}b displays this re-scaling of the critical values. For tree-like networks, such as 2-clique networks, $R_\mathrm{e} = 1$ corresponds to a constant (horizontal line) in the re-scaled representation, while for networks with cliques, this value increases with $\alpha$. This indicates that networks with cliques require much larger effective transmission probabilities to reach the epidemic threshold compared to what is expected by the tree-like approximation, with the difference growing as the isolation probability $\alpha$ increases. This shows the moderating effect that clique structures can have on an epidemic in the presence of contact tracing, as it helps to cut not only onward infections but also local indirect spreading paths in the network.

\begin{figure}
\includegraphics[width=.495\linewidth]{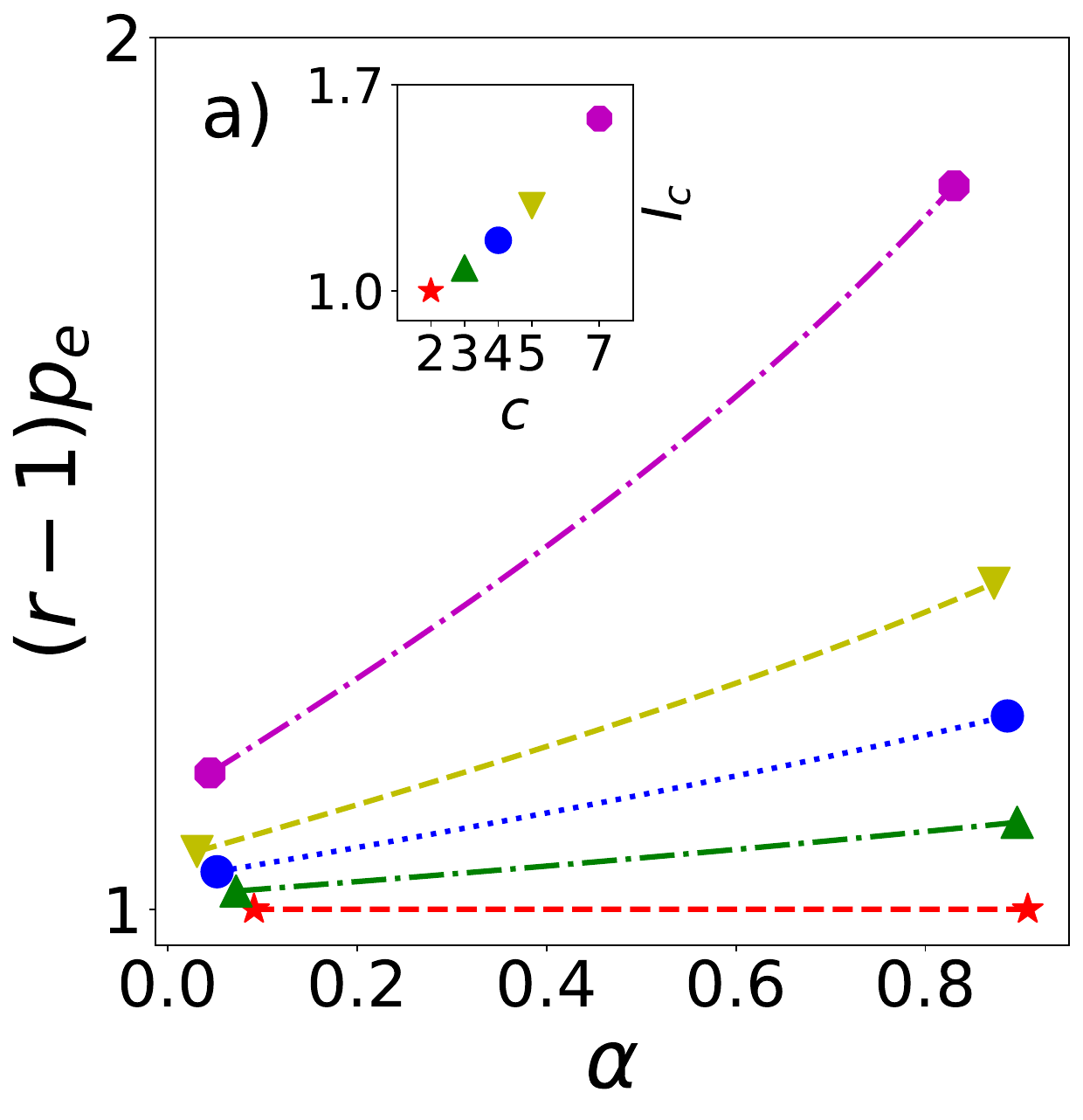}%
\includegraphics[width=.505\linewidth]{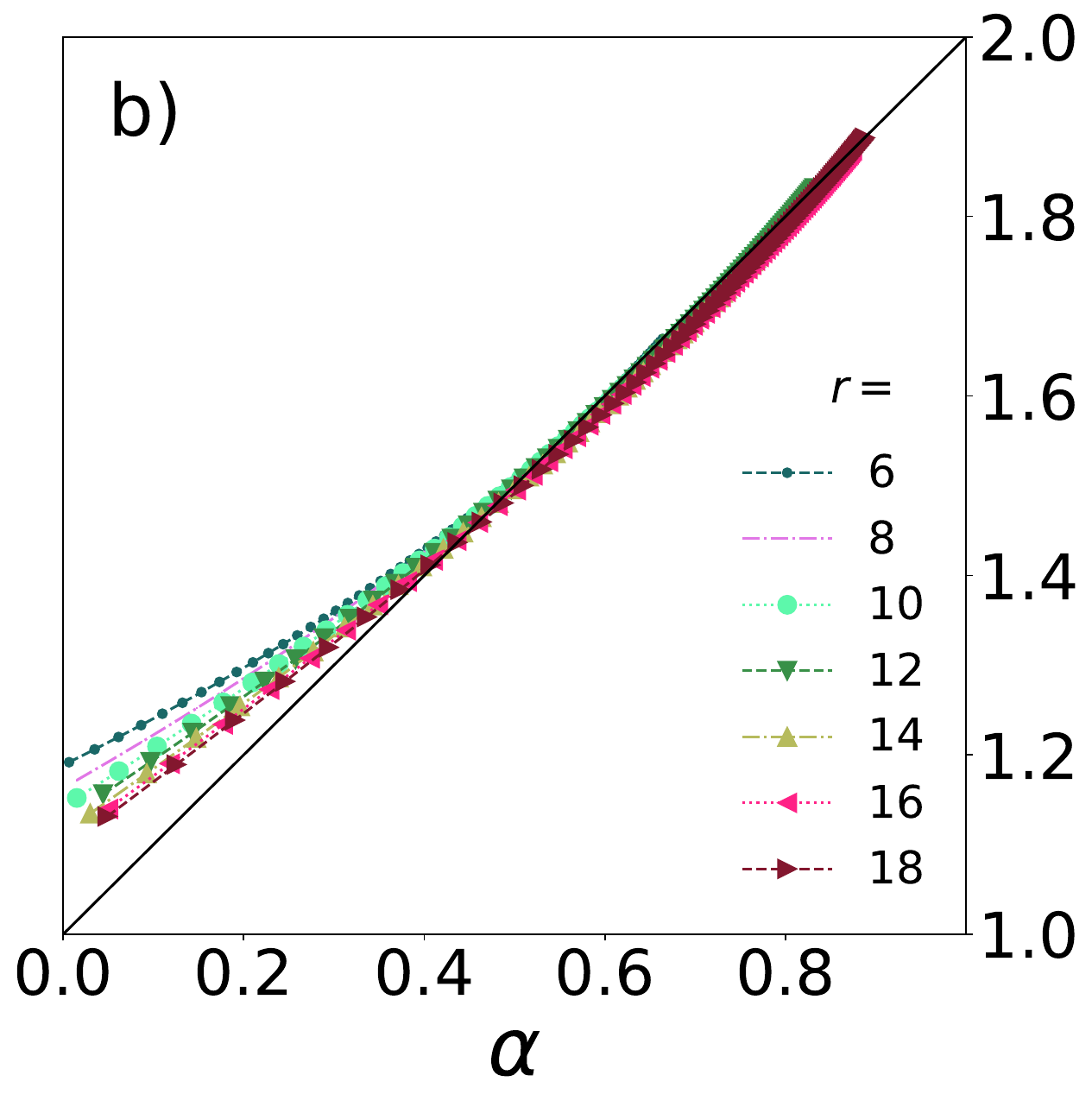}
    \caption{Critical curves re-scaled as effective branching factors for tree-like networks $(r-1)p_\mathrm{e}$. Panel (a) is the same as panel Fig.~\ref{fig:critical_curve_theory}b, but on the vertical axis, the effective transmission probabilities are multiplied by the excess degree. Further, results are shown for a larger network where $r = 12$. The red curves marked with stars are for the case that there $c=2$ and the network is tree-like, therefore Eq.~\ref{eq:critical} holds such that $R_\mathrm{e} = p_{\mathrm{e}} \bar{d} = p_{\mathrm{e}} (r-1) = 1$ for any $\alpha$ value. For $c>2$, this equation does not hold. The right end markers on panel a indicate extreme points from Eq.~\ref{eq:extremey}. The inset shows the relative maximum increase in the effective epidemic threshold for different networks with cliques. Each point in the inset is the ratio of the $p_\mathrm{e}$ values at the endpoints of each curve outside the inset, such that $I_c = p_\mathrm{e}(\alpha_\mathrm{max})/p_\mathrm{e}(\alpha_\mathrm{min}) $. (b) This panel shows how $(1-r)p_{\mathrm{e}}$ changes when we have networks with different degrees, $r = 6$ to $r=18$, and maximal connectivity-preserving clique size (i.e., when $n_c=2$).  }
\label{fig:critical_degrees}
\end{figure}

The transition points strongly depend on the node degree, $r$, in the class of clique networks we consider here. For most of the analysis in this paper, we kept the node degree fixed to $r=6$ (Fig.~\ref{fig:critical_curve_theory}); here, we turn our attention to how node degree can affect disease spreading. We do this by varying $r$ and holding other network properties constant. This is interesting because the node degree, $r$, rescales the transition points between sub- and super-critical regions. More precisely, if we use the excess degree to scale the critical $p_\mathrm{e}$ by plotting $(r-1)p_\mathrm{e}$ as a function of $\alpha$, the phase diagram returns to a scale that is independent of $r$ such that the $c=2$ line is precisely at $(r-1)p_\mathrm{e}=1$. This rescaling is illustrated for $r=6$ and $r=12$ in Fig.~\ref{fig:critical_curve_theory} and Fig.~\ref{fig:critical_degrees}a for valid network configurations, respectively. Recall that the networks we study still must obey $n_c(c - 1) = r$.

A more systematic exploration can be found in Fig.~\ref{fig:critical_degrees}b, where each node belongs only to two cliques, $n_c=2$, representing the extreme non-trivial scenario where each clique is as large as possible without the network consisting solely of isolated cliques. In this case, the critical curves collapse in the re-scaled plot for a range of $r=6$ to $r=18$ that we tested. The collapse approximately follows a straight line from $(r-1)p_\mathrm{e}=1$ for $\alpha = 0$ to $(r-1)p_\mathrm{e}=2$ for $\alpha=1$, with the approximation getting better for larger values of $\alpha$. Note that for $\alpha = 0$, the critical transmission probability equals the critical bond percolation probability for the SIR model \cite{newman2002spread, stegehuis2021network, ziff2021percolation, valdez2023epidemic}. 

\begin{figure}
    \includegraphics[width=.6\linewidth]{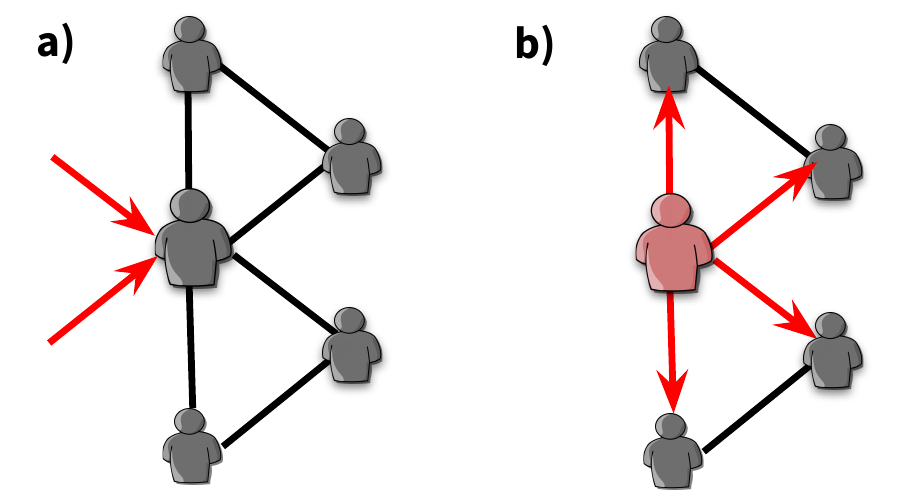}
    \caption{Spreading process in a $6$-regular $3$-clique network under the extreme case of $p=1$, where the infection propagates severely through the cliques. In this scenario, (a) an active node (infected but not in quarantine) can transmit the disease to other nodes in its adjacent cliques, (b) resulting in new infections. The number of new infections caused by an active node is not equal to the total number of members in the cliques that the node belongs to, which is $n_c(c-1)$, but rather the number of members in the cliques attached to that node, excluding the clique that infection is coming from, which is $(n_c -1)(c-1) = 2 \times 2$. For further details on the network structure and spreading dynamics, refer to Sec.~\ref{sec:Networks} and \ref{sec:SIRQ}. }
\label{fig:p=1}
\end{figure}

The curve collapse can be understood by examining the extreme case  $p=1$ (where the infection always succeeds), and the critical point for contact tracing probability $\alpha_*$ (i.e., how large does $\alpha$ need to be to prevent an outbreak when $p=1$). In this case, infected nodes always infect all of their neighbors, and during the early stages of the epidemic, each clique either has (1) exactly one infected node and $c-1$ susceptible nodes, (2) one recovered node and $c-1$ infected nodes (out of which a fraction of $\alpha$ are isolated in expectation), or (3) only susceptible nodes. When an infection arrives at a clique, the infected node, which is not in quarantine, can spread the infection to $(n_c - 1)(c-1)$ new nodes (offspring) in the next time step, as illustrated in Fig.~\ref{fig:p=1}. Therefore, $\alpha_*$ can be obtained by setting the expected number of active (infected but not in quarantine) nodes to one, which occurs when $(1-\alpha)(n_c - 1)(c-1) = 1$. Therefore, the critical value for $\alpha$ is given by
\begin{equation}
    \alpha_{*} = 1 - \frac{1}{(n_c - 1)(c-1)}.
    \label{eq:extremeP=1}
\end{equation}
Markers in the right end of Fig.~\ref{fig:critical_curve_theory}b shows such extreme points $(\alpha_{*}, 1-\alpha_{*})$.
Substituting $\alpha_{*}$ into $(r-1)p_{\mathrm{e}} = (r - 1)(1-\alpha)$ and recalling that $r = n_c(c-1)$, yields 
\begin{equation}
    (r-1)p_\mathrm{e} = \frac{n_c(c-1) -1}{(n_c -1)(c-1)}.
    \label{eq:extremey}
\end{equation}
Markers in the right end of Fig.~\ref{fig:critical_degrees}a shows such extreme points $\big(\alpha_{*},  (r-1)(1-\alpha_{*})\big)$. When the clique size $c$ grows to infinity, Eq.~\ref{eq:extremey} leads to
\begin{equation}
    (r-1)p_\mathrm{e}\xrightarrow[]{c\rightarrow \infty } \frac{n_c}{n_c - 1},
    \label{eq:extreme_c}
\end{equation}
which yields the value $2$ when $n_c=2$, for example. This explains why the re-scaled critical infection probability curves approach $2$ when $\alpha$ is sufficiently large, as seen in Fig.~\ref{fig:critical_degrees}b.

Using our multitype branching process description of contact tracking, not only can we unpick the interplay of clique structure on the criticality of the process, which we have explored in this section, but we can also estimate the expected outbreak size. This is the focus of the following section. But before proceeding, it is worth noting that integrating cliques into network models increases the clustering coefficient. However, it is essential to recognize that increasing clustering in networks can be done in different ways, and it may lead to changes in other network properties, such as degree heterogeneity. Therefore, since clustering alone does not solely dictate the epidemic threshold \cite{newman2009random, karrer2010random, miller2009percolation, gleeson2010clustering, ball2010analysis, miller2009spread, melnik2011unreasonable,  coupechoux2014clustering,  van2015hierarchical, stegehuis2016epidemic, mann2021random, stegehuis2021network} or the component size distribution \cite{ritchie2016beyond} of network, running our SIRQ dynamics on any clustered networks may not necessarily lead to the same results we obtained here.
 
\subsection{Sub-critical Outbreak Sizes}
\label{sec:Outbreak_Size_Revisited}
Characterizing the spreading process described above also provides access to methods for calculating quantities of interest, such as the epidemic size, via the next-generation matrix. We are interested in the outbreak size (the expected total number of infected individuals in an outbreak) for a given parameter set. We follow closely the method outlined in Ref.~\cite{keating2022multitype}, where they derived the expected epidemic size, $E$, in the sub-critical regime. We consider the contributions for subtrees seed of each motif type ($\vec{z}\,$) --- as well as the expected number of offspring of each type from each motif type, which has already been discussed in Sec.~\ref{sec:m-field} via the next-generation matrix. We also need to consider the number of infected nodes contributed by each type, which will be given by the vector $\vec{I}$. The expected epidemic size, $E$, can be found by solving the following two equations:  

\begin{equation}
    \left( \mathbb{I} -\textbf{M}^{T} \right)\vec{z}=\vec{I}\label{eq:x_inf}
\end{equation}
and
\begin{equation}
    E = 1 +(\vec{z}^{\,0})^{\,T}\textbf{M}^{\,T}\vec{z}, \label{eq:X_inf}
\end{equation}
where $\vec{z}^{\,0}$ is the initial seeding of each motif type at the start of the process, and $\mathbb{I}$ is the identity matrix. For example, consider the case where we have each node as being part of three cliques, where each clique contains three nodes (see Fig.~\ref{fig:clique_states}), the elements of $\vec{z}^{\,0}$ are $(3, 0, 0, 0)^T$ and $\vec{I}$ are $(0, 1, 2, 1)^T$. The first element of $\vec{z}^{\,0}$ is $3$, as each node is a member of three cliques, and, as this is the seed configuration, the contagion has not spread to any other nodes, leaving all the other motif types zero. We simply count the number of active nodes in each motif for the elements of $\vec{I}$. Referring again to Fig.~\ref{fig:clique_states}, types $Z_2$ and $Z_4$ both have one active node, and type $Z_3$ has two active nodes. We need to be careful not to double-count nodes, and as such, we set the first element of $\vec{I}$ to zero. For the full derivation, please refer to Ref.~\cite{keating2022multitype}. Using this, we can now easily sweep through a parameter set to find the relationship between $p$, the initial probability of infection, and $\alpha$, the probability of quarantine on the expected epidemic size under our mean-field view of the disease process.

\begin{figure}
\includegraphics[width=\linewidth]{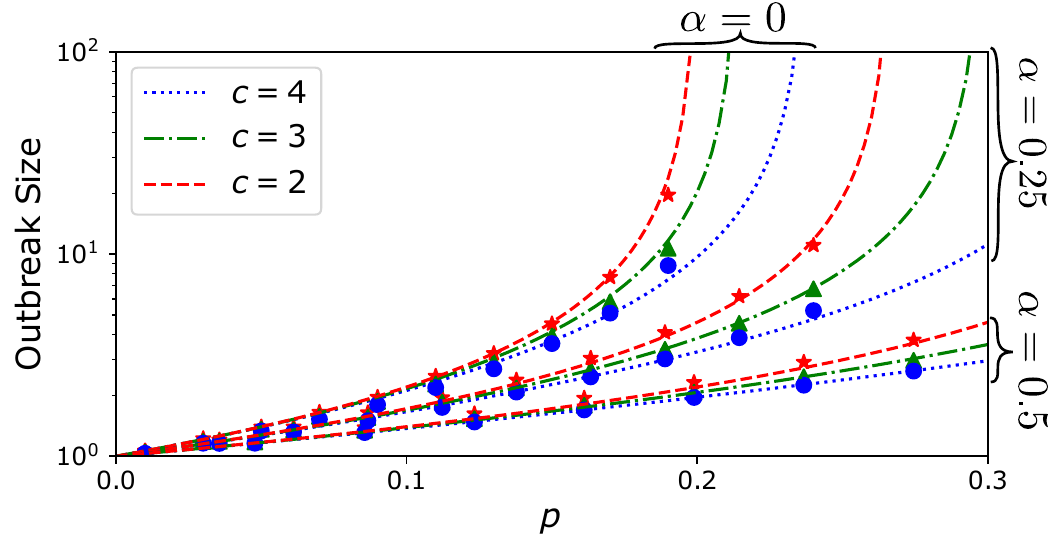}
     \caption{Outbreak size in the sub-critical regime as a function of $p$ across the three network structures while considering the influence of isolation probability, $\alpha$. Nine curves are grouped into three sets (from left to right) according to their isolation probabilities, with each set containing three curves with $\alpha = 0.0, \alpha = 0.25,$ and $\alpha = 0.5$. These groupings demonstrate the influence of $\alpha$ on outbreak size in the sub-critical regime. Increasing $\alpha$ or clique size reduces the outbreak size, as calculated by Eq.~\ref{eq:x_inf} and Eq.~\ref{eq:X_inf} using the next-generation matrix from mean-field approximation found in Sec.~\ref{sec:Outbreak_Size_Revisited}. Markers represent the results of 50,000 simulations, while dotted lines depict the results of the mean-field calculations presented in Sec.~\ref{sec:SIR}.}
    \label{fig:expected_epi_size}
\end{figure}

Looking at the qualitative behavior of the curves, we see from Fig.~\ref{fig:expected_epi_size} that as we increase $p$, naturally, the expected epidemic size rises; however, when we increase $\alpha$, we see that the average size of the outbreak decreases across all three network topologies that we consider. Moreover, this effect is most pronounced for networks with larger cliques, as this network gives the quarantining behavior more opportunities to remove possible infection paths via our mean-field approximation. 
Note that the outbreak size in the sub-critical regime does not scale with the network size and decreases dramatically by increasing $\alpha$.

In the next section, we will present a complex contagion approximation to contact tracing for calculating the epidemic thresholds. The results of  Fig.~\ref{fig:expected_epi_size}  also hold for this approximation to the model, which is discussed in the next section. Please refer to Appendix \ref{sec:exp_cas_size_cc} for the complex contagion approximation calculation for the expected cascade size.  

\section{Complex Contagion Formulation}
\label{sec:Complex_contagion}
Next, we will show that the model described in Sec.~\ref{sec:SIR} is closely related to a SIR model, which allows the probability of infection to change as a function of infection attempts. In this related complex contagion model, we do not keep track of isolated node states. Instead, we keep track of failed infection attempts on the susceptible nodes. Since we know that a contact tracing attempt preceded each infection attempt, we know there has been an equal number of infection and contact tracing attempts. In this sense, each infection attempt also carries a risk of isolation, and every attempt becomes less likely to yield an infected node. This is in contrast to the typical social complex contagion processes where each infection attempt makes it more likely for the next one to succeed.

We borrow the framework of the multitype branching processes for complex contagion as introduced in Ref.~\cite{keating2022multitype}, where the infection state of the nodes characterizes motifs. Notably, the model is the same apart from $p_k$ as has been previously defined in Ref.~\cite{keating2022multitype}, where the probability of adoption after $k$ attempts is given by 
\begin{equation}
    p_k=1- (1-p) (1-\alpha)^{k-1}\,.
    \label{eq:pk_social}
\end{equation} 
That is, the $\alpha$ parameter works exactly opposite to the contact tracing here: the larger the $\alpha$ value, the larger the probability that infection attempts beyond the first one are likely to succeed.

\begin{table}[b]
\begin{tabular}{ c c c } 
 \hline
 \hline
$i,j$ & \hspace{3mm}&$m_{ij}$ \\
\hline
$ 1 , 1 $&&$ 4 p \left(1 -\alpha \right)  $\\
$ 1 , 2 $&&$ 2p \left(1-\alpha \right)^{2}  $\\
$ 2 , 1 $&&$ 2 p \left(1-\alpha \right) \left(1- p \left(1-\alpha \right) \right) $\\
$ 3 , 1 $&&$ p^{2} \left(1-\alpha \right)^{2} $\\
$ 4 , 2 $&&$ p \left(1-\alpha \right)^{2} $\\
 \hline
 \hline
\end{tabular}
\caption{Non-zero elements of the next-generation matrix $\mathbf{M}_{4\times 4}$ in the complex contagion approximation for a 3-clique network. $m_{ij}$ gives the expected number of $Z_i$ cliques from a $Z_j$ clique. As shown in Fig.~\ref{fig:clique_states}.}
\label{table:m44}
\end{table}

In the SIR process with contact tracing
each isolation fails with probability $1- \alpha$ and a node is not isolated after $k$ attempts with probability $(1-\alpha)^k$. If the person is not isolated, they have a probability of $p$ being infected by an adjacent infected node. In total, the probability that a node gets infected by a neighbor after $k$ attempts, given that it is not yet infected, is
\begin{equation}
    \hat{p}_k=p (1-\alpha)^k\,.
    \label{eq:pk}
\end{equation}
This probability describes exactly the opposite behavior to typical social complex contagion processes, where the probability of infection increases with the number of attempts. Suppose we do not track whether the susceptible node is quarantined or the infection has failed even though the node was not quarantined. In that case, we can follow the method and formulas given in Ref.~\cite{keating2022multitype} by simply replacing the probability $p_k$ of Eq.~\ref{eq:pk_social} with $\hat{p}_k$ from Eq.~\ref{eq:pk}. 
In this picture, isolated and susceptible nodes are treated the same and put into the susceptible compartment. We are not explicitly tracking individuals in the $\mathrm{Q}$ compartment. However, we have made the probability of infection a function of the number of exposures as given by Eq.~\ref{eq:pk}. In this approximation, we also retain the infected and quarantined nodes $\mathrm{Q_I}$ in the susceptible compartment $\mathrm{S}$, which means that a $Z_1$ cannot make a $Z_4$ directly (Fig.~\ref{fig:clique_states}).

\begin{figure}
\includegraphics[width=.51\linewidth]{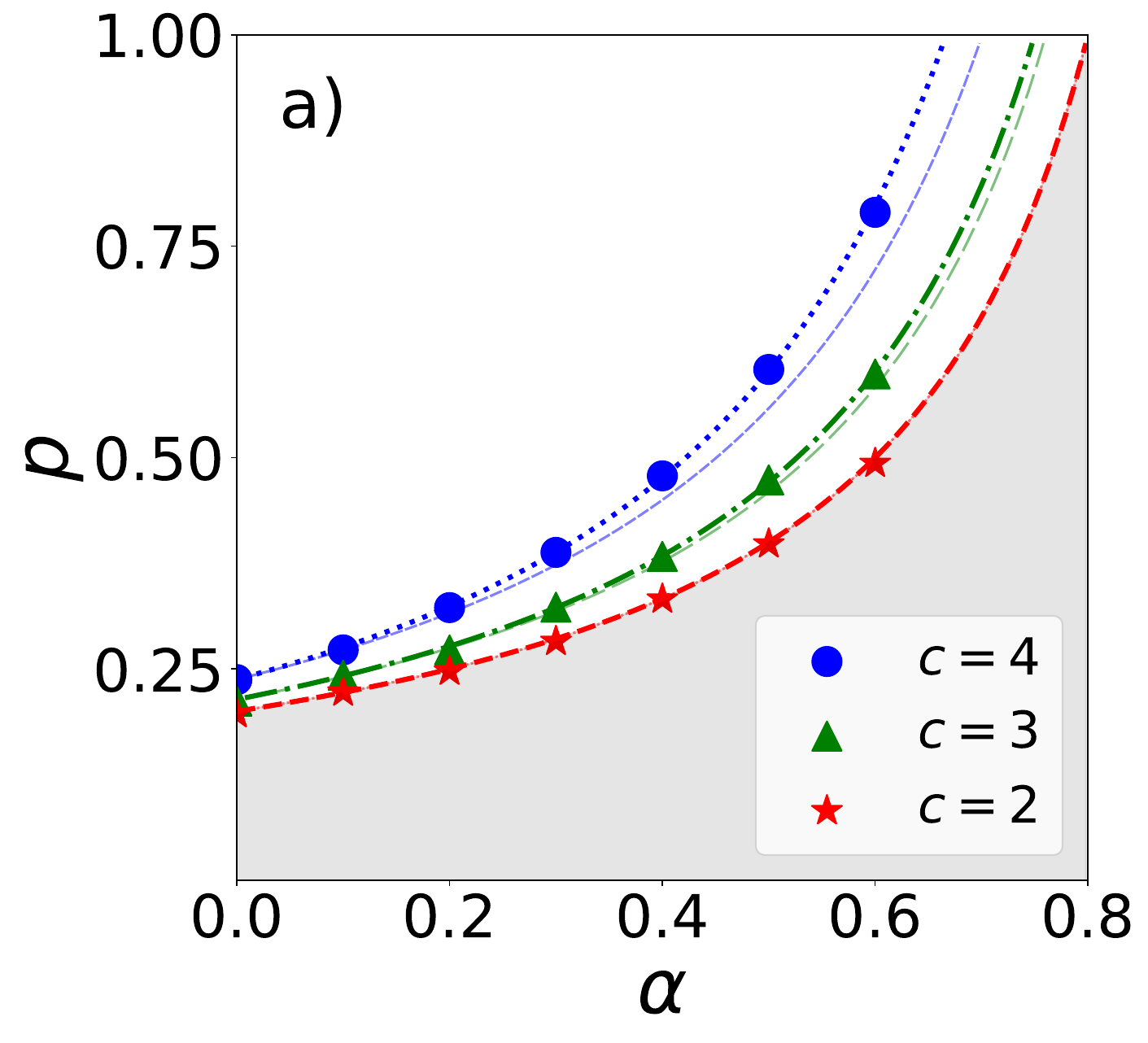}%
\includegraphics[width=.49\linewidth]{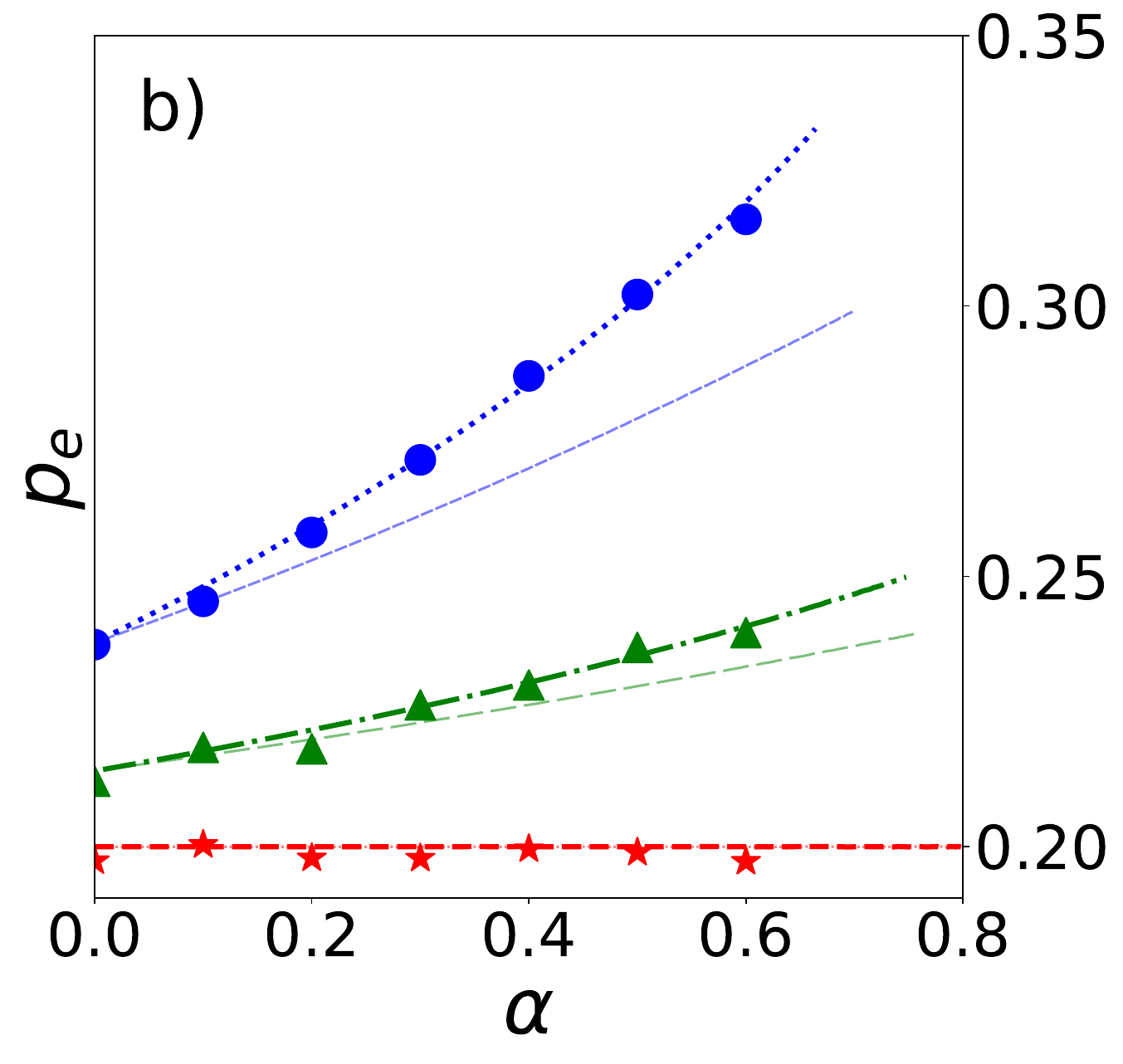}
    \caption{Comparison of the mean-field approximation introduced in section Sec.~\ref{sec:SIRQ} with the complex contagion approximation of Sec.~\ref{sec:Complex_contagion}. Panels are similar to the panels of Fig.~\ref{fig:critical_curve_theory}. Here, markers are the results of the simulations described in Sec.~\ref{sec:SIR}. 
    Our simulation results align with the bold curves, which are the results of the mean-field approximations of Sec.~\ref{sec:R0}. The thin curves result from the complex contagion approximation introduced in Sec.~\ref{sec:Complex_contagion}, which deviates slightly more from the other mean-field approximation when $\alpha$ increases. The complex contagion approximation gives the lower bound of critical transmission probability for all values of $\alpha$.}
\label{fig:critical_curve_theory2}
\end{figure}

Table~\ref{table:m44} shows the non-zero elements of the next-generation matrix for a 3-clique network, and Fig.~\ref{fig:critical_curve_theory2} shows the result of the complex contagion approximation when we set the spectral radius of the new next-generation matrix to unity. The results of this approximation are close to the simulations and the mean-field solution presented in Sec.~\ref{sec:R0} when the quarantine probability $\alpha$ and clique size $c$ are small. Larger $\alpha$ and $c$ values will lead to underestimated epidemic threshold values for the infection probability $p$. As the critical curves of the complex contagion are positioned below those of the full contact tracing (see Fig.~\ref{fig:critical_curve_theory2}), it establishes the complex contagion process as the lower bound (or estimate) for the entire contact tracing process. With the complex contagion formulation, the advancement in epidemic thresholds is less pronounced. This is due to the slower upward trajectory of the curves with increasing alpha, contrasting with the original contact tracing formulation. So, the phenomenon of disease spreading under contact tracing in networks with cliques can be understood to be analogous to social complex contagion but with the opposite and more minor effect. Intuitively, analogously to social complex contagion, this explains why it is crucial to consider contact tracing in network models that contain realistic group structures.

\section{Conclusion \& Discussion}
\label{sec:conc}
In this work, we incorporate contact tracing into disease-spread models on social networks, focusing on how local group structures, modeled with cliques, impact the epidemic thresholds and sizes. This model, contrasting with traditional assumptions of fully mixed or tree-like networks, demonstrates greater efficacy of contact tracing in networks with clustering. Moreover, we show that disregarding group structure in contact tracing is analogous to ignoring group structure in complex contagion models where previous exposures increase the chance of adoption/infection. This illustrates the possible benefits of contact tracing in real-world settings, especially in the early stage of disease spread, where quarantining limits the possible paths of infection that disease can take through a network. 

The dynamical model we used in this paper is an idealized representation and, therefore, oversimplifies the complexities of disease transmission and contact tracing. In actual social situations, the implementation of contact tracing may vary across different groups within the network, resulting in strong effects on infection risk and threshold size. 
Further, our model integrates factors related to contact tracing and isolation timing into a single parameter $\alpha$. Contact tracing is always either entirely successful or unsuccessful in this simplification. In reality, contact tracing could be partially successful such that the isolated individual passes on the infection to part of the contacts that would have been infected without any intervention.
Additionally, there could be additional effects in networks with cliques that are affected by the timing. Further research is needed to understand how quarantining measures impact epidemics on more realistic contact networks with contact tracing and how outbreak sizes are distributed in such settings.

The details of the epidemic model itself could also impact the effectiveness of contact tracing in networks with cliques. For example, asymptomatic individuals can be crucial in disease spread as they can unknowingly transmit the infection and therefore do not lead those people being contact traced. However, in the presence of social groups, both the asymptomatic individuals and the people infected by them can be isolated through indirect connections, potentially alleviating the problems caused by asymptomatic individuals. Similarly, one could incorporate a non-contagious \textit{exposed} phase using the SEIR model \cite{newman2002spread, kenah2011epidemic} or various other complications that would make the models more realistic, such as temporal inhomogeneities of the contact networks \cite{Saramaki2023, badie2022directed, badie2022directed2}. 

Our model's strength lies in its simplicity and general modeling practicality. It encapsulates identifying infected individuals, alerting their contacts, and isolating those potentially exposed within a single parameter, denoted $\alpha$. This versatile model can be applied to various interventions akin to contact tracing. For instance, social distancing could theoretically fit into this model if we consider exposed individuals maintaining a large enough physical distance with probability $\alpha$ from all their peers, thus mirroring the concept of self-quarantine. While social distancing may be regarded as a form of partial or complete self-isolation, the essence of both practices—and of quarantining—is fundamentally similar. Although these public health terms are often finely distinguished in specific contexts, the underlying principles governing these interventions are consistent.

In summary, our results highlight the importance of considering realistic social network structures when modeling epidemics and interventions. Our model is deliberately simplistic and is used to isolate key insights. The key conclusions we draw are: (1) contact tracing is more efficient in social networks with groups than one would expect based on the tree-like models, (2) the effect of groups is more prominent if the groups are larger, and if the contact tracing is more efficient, and finally (3) SIR spreading under contact tracing can be approximately understood as a complex contagion process where multiple exposures reduce the infection probabilities.

\section*{Code Availability}
This study's simulations and numerical computations are publicly available at \cite{simulations}.

\section*{Acknowledgement}
A. K.~R. would like to thank Lasse Leskelä and Takayuki Hiraoka for the fruitful discussion during the preparation of this work. The simulations presented above were performed using computer resources within the Aalto University School of Science ``Science-IT'' project.  A. K.~R. and M. K. acknowledge funding from the project 105572 NordicMathCovid which is part of the Nordic Programme on Health and Welfare funded by NordForsk. This work was also supported by the Academy of Finland (349366, 353799) and by  Science Foundation Ireland [grant numbers 18/CRT/6049 (L.A.K.), 16/IA/4470 (J.P.G.), 16/RC/3918 (J.P.G.), 12/RC/2289 P2 (J.P.G.)] with co-funding from the European Regional Development Fund. For the purpose of Open Access, the author has applied a CC BY public copyright license to any Author Accepted Manuscript version arising from this submission.

\bibliography{citations}

\appendix

\section{Expected epidemic size for the complex-contagion process} \label{sec:exp_cas_size_cc}
\begin{figure}
 \includegraphics[width=\linewidth]{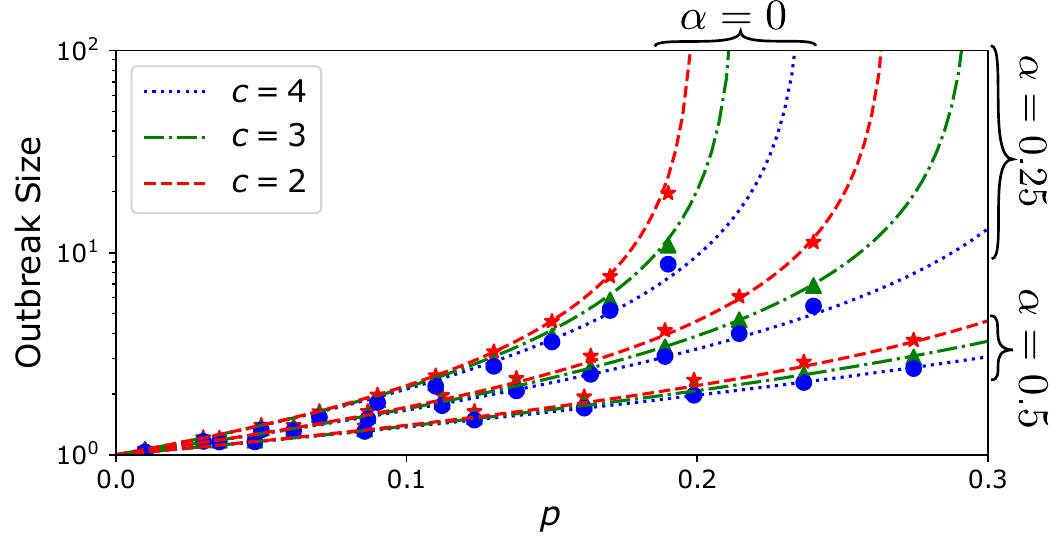}
     \caption{Outbreak size in the sub-critical regime as a function of $p$ across the three network structures while considering the influence of isolation probability $\alpha$. Nine curves are grouped into three sets (from left to right) according to their isolation probabilities, with each set containing $\alpha = 0, \alpha = 0.25,$ and $\alpha = 0.5$. These groupings demonstrate the influence of $\alpha$ on outbreak size in the sub-critical regime. Increasing $\alpha$ or clique size reduces the outbreak size, as calculated by Eq.\ref{eq:x_inf} and Eq.\ref{eq:X_inf} using the next-generation matrix from mean-field approximation found in Sec. \ref{sec:Complex_contagion}. Markers represent the results of 50,000 simulations, while dotted lines depict the results of the mean-field calculations presented in Sec.~\ref{sec:Outbreak_Size_Revisited}.}
    \label{fig:expected_epi_size_cc}
\end{figure}

We can calculate the expected epidemic size for a multi-type branching process in the sub-critical regime, as explained in Sec.~\ref{sec:Outbreak_Size_Revisited}. Here, we examine the expected epidemic sizes for the complex contagion process. To do this, we need only to replace the next-generation matrix from the mean-field description in Sec.~\ref{sec:m-field} with that of the complex-contagion described by Eqs. \ref{eq:x_inf} and \ref{eq:X_inf}.
As with the results from Sec.~\ref{sec:Outbreak_Size_Revisited}, we find good agreement when comparing the resulting curves for the expected epidemic (cascade) size, $E$, to simulation results in Fig. \ref{fig:expected_epi_size_cc}. When examining the qualitative behavior of the curves for expected epidemic size, we see from Fig. \ref{fig:expected_epi_size_cc} that as we increase $p$, as expected, the average epidemic size increases. However, increasing $\alpha$ decreases each network's average outbreak size. Moreover, this effect is most pronounced for networks with larger cliques, as this network gives the quarantining behavior more opportunities to remove possible infection paths via the complex-contagion approximation to the process.

\section{Effective Reproduction Number}
\label{sec:r0_old}
We investigate the role of contact tracing in networks with cliques and its effect on the effective reproduction number, $R_\mathrm{e}$. Our analysis reveals that $R_\mathrm{e}$ reduces in response to increases in the tracing probability, $\alpha$, and variations in the transmission probability, $p$. To find $R_\mathrm{e}$ in our simulations, we run our discrete-time dynamics and count, on average, how many people have been infected in each time step by a typical infected node. In different trials, we start with a single infected node, chosen uniformly randomly in the network. We follow how many susceptible nodes it infects, even counting the isolated ones, such as the case in Fig.~\ref{fig:demo}a. For finite networks, this means that we need to run the simulation long enough for the process to stabilize but short enough that the ratio of the infected nodes to the network size remains close to zero. If no new infection happens in a generation, the disease has died out. 
The total number of new infections caused by the seed node would be the \textit{individual} reproduction number of that node, and we report it as the effective reproduction number of generation $t=1$. In the next step, we do the same for the resultant active (infected but not in quarantine) nodes generated by the seed node, one-after-another. Notice that from this step on, a neighbor of an active node may be in quarantine because of its interaction with other neighbors, not with the one it is receiving the infection from, such as the case shown in Fig.~\ref{fig:demo}b. In these cases, the active node cannot infect the node that has been in quarantine via other nodes. When we are done with all the active nodes of this generation, we report the average number of individual reproduction numbers of these nodes as the effective reproduction at generation $t=2$. The process can be continued for more generations, depending on the network size. For large enough networks, the ensemble average of effective reproduction number over different trials starts to stabilize from generation $t =3$ in the parameter ranges we have explored, indicating that its value remains constant for some time, depending on the network's size. Hence, we consider this stabilized value at $t=3$ as the effective reproduction number $R_\mathrm{e}$.

\begin{figure}
\includegraphics[width=.3662\linewidth]{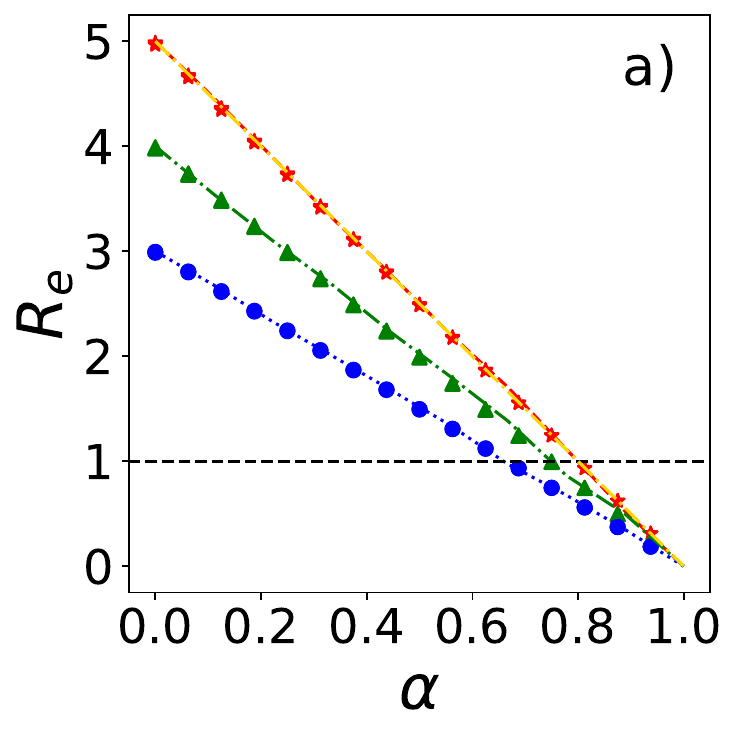}%
\includegraphics[width=.3151\linewidth]{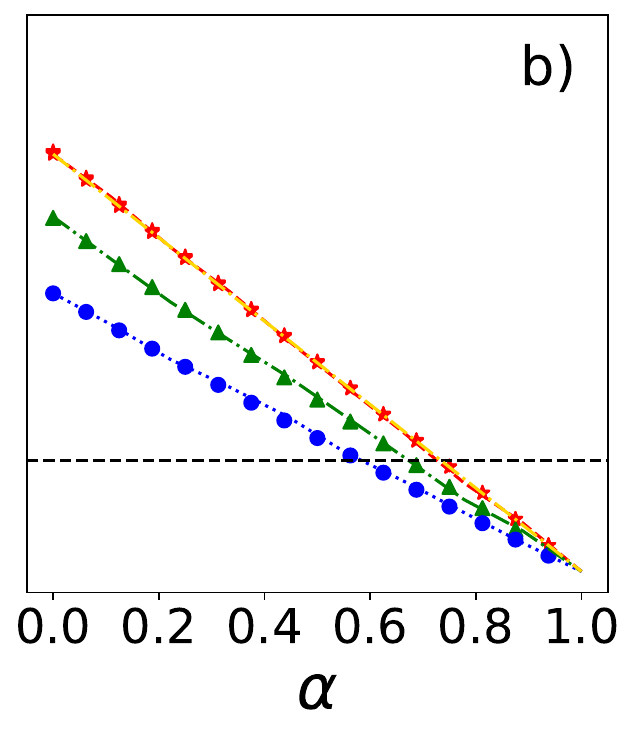}%
\includegraphics[width=.3151\linewidth]{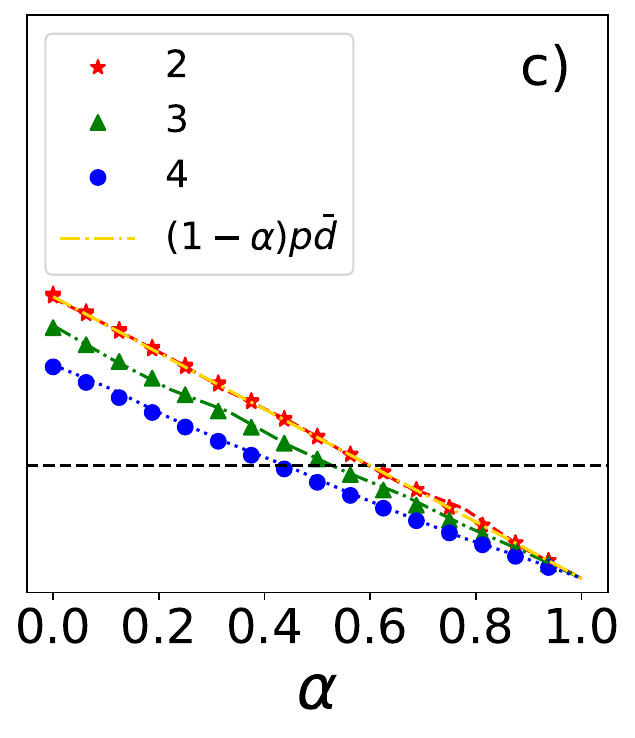}
\caption{How contact tracing on clique network reduces the effective reproduction number $R_\mathrm{e}$ by increasing $\alpha$. Effective reproduction number for different networks with cliques with $r=6$ for transmission probabilities (a) $p = 1.0$, (b) $p = 0.75$ and (c) $p =0.5$. The dotted lines are from the mean-filed calculations introduced in Sec.~\ref{sec:R0}, and the markers are from Monte Carlo simulations described in Sec.~\ref{sec:SIR}. 
The yellow dashed lines, which overlap with the red curves with stars, represent Eq.~\ref{eq:critical}, which match the cases where $c=2$ (tree-like networks). The larger the transmission probability, the larger the differences between the curves of different networks with cliques. The larger the clique size, the less effort we need to quarantine people since networks with larger cliques reach $R_\mathrm{e} = 1$ for lower values of $\alpha$.}
\label{fig:r0_old}
\end{figure}

We observe that in networks with cliques of size $r=6$ and $c=2,3,4$, and transmission probabilities of $p=1, 0.75, 0.5$, larger cliques correspond to a reduced need for quarantine measures to bring $R_\mathrm{e}$ down to 1. This relationship between clique size and the required intensity of quarantine efforts to control the spread is illustrated in Fig.~\ref{fig:r0_old}, which aligns with our theoretical predictions of Sec.~\ref{sec:R0} and simulations described in Sec.~\ref{sec:SIR}.

\section{Reproduction Number and the Next-generation Matrix}
\label{sec:irreducibility}
We will next give additional details on using the leading eigenvalue of the $\mathbf{M}$ matrix as the reproduction number $R_\mathrm{e}$. Two specific issues were only briefly discussed in the main text: why does the leading eigenvalue of the reducible matrix $\mathbf{M}$ tell us about the critical behavior, and why can the leading eigenvalue be interpreted as $R_\mathrm{e}$?

The argument for the leading eigenvalue of the reducible matrix $\mathbf{M}$ determining the criticality is very similar to the one made in Sec.~IV~A~2 of Ref.~\cite{keating2022multitype}. 
We can divide the transitions between $Z$ states into two categories: ones that describe changes within a clique and ones that describe the infection arriving in a clique that previously only had susceptible nodes. The transitions within cliques form a directed acyclic graph (DAG), because the SIRQ process always moves in one direction, i.e., from susceptible to infected or removed or from infected to removed. DAGs can always be put into upper (or lower) triangular form.  
Every motif that has at least one susceptible and one infected node will have a transition to the $Z_1$ state, which has one infected node in an otherwise susceptible clique. We can permute the matrix $\mathbf{M}$ by collecting these motifs to a block $\mathbf{B}_1$. These motifs form a strongly connected block because $Z_1$ is the root of the DAG, and the block is thus irreducible. The remaining states are still triangular, where each motif forms its own block $\mathbf{B}_i$, so in total, we have the matrix in a normal form. Further, the motifs that do not belong to the block $\mathbf{B}_1$ are dead ends as they cannot produce any offspring with infected nodes.

The method described above can always be used to write the reducible matrix $\mathbf{M}$ in its normal form such that blocks $\mathbf{B}_i$ fill the upper triangle part of it. Since $\mathbf{M}$ is non-negative, the spectrum of  $\mathbf{M}$  is the union of the spectra of the $\mathbf{B}_i$ \cite{varga1999matrix}. Here the ${B}_i$ has zero eigenvalues for $i \neq 1$, and the largest eigenvalue of $\mathbf{M}$, i.e., the Perron root, is the same as the largest eigenvalue of $\mathbf{B}_1$ which is an irreducible matrix. Given that we initialize our spreading process sparsely such that there are only $Z_1$ motifs in the network in the beginning (in addition to the fully susceptible ones we do not track), the long-term dynamics will always be governed by the leading eigenvalue of $\mathbf{B}_1$ (and therefore $\mathbf{M}$)which \cite{keating2022multitype}.

The $R_\mathrm{e}$ correspondence to the leading eigenvalue might initially seem non-intuitive, considering that some transitions create multiple infected nodes at one step. This indicates that one needs to multiply the effects of transitions by the number of newly infected nodes in them to compute the expected number of newly infected nodes a typical infected node produces. However, this is not necessary. A key observation here is that the number of $Z_1$ motifs is directly proportional to the number of infected nodes in the network. Every time a node is infected, it will create $n_c-1$ of new $Z_1$ motifs, where $n_c$ is the number of cliques each node belongs to. In the next time step, those motifs transition into one of the other motif types, and the infected individuals become removed, so the number of $Z_1$ motifs is always updated to be $n_c-1$ times the number of infected nodes. That is, at time step $t$, the number of infected nodes is $I^t = z_1^t / (n_c-1)$, where $z_1^t$ is the number of motifs $z_1$ at time $t$. Given that we are at the steady state, $z^t$ is the leading eigenvector, i.e., $z^{t+1} = M z^t = \lambda  z^t$, and  $z_1^{t+1}= \lambda  z_1^t$, which means that $I^{t+1} = \lambda  I^t$.

\section{Automatic Generation of the Next-generation Matrices}
\label{sec:auto_matrices}

\begin{figure}[h!]
\centering
\includegraphics[width=.9\linewidth]{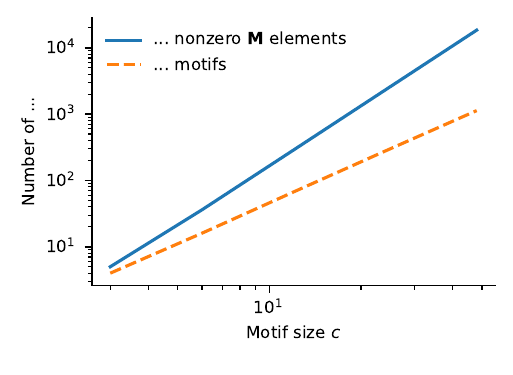}
\caption{Number of motifs and the number of nonzero elements of $\mathbf{M}$ for given clique size $c$. 
}
\label{fig:Number of motifs}
\end{figure}

We show how the next-generation matrix can be constructed for any clique size $c$ by an algorithm described here. The construction is done by going through every motif and considering the transitions and expected values leading out of that motif. That is, we go one motif at a time, starting from the one with only a single infected node and the rest of the nodes susceptible, constructing the motifs one time and stepping away from that motif and the expected number of new motifs produced.

We can note that each clique motif is uniquely defined by the number of nodes in each compartment; that is, we define each unique clique motif by $Z_i=(n_{\mathrm{S}},n_{\mathrm{I}},n_{\mathrm{R}})$ as one that has $n_{\mathrm{S}}$ susceptible nodes, $n_{\mathrm{I}}$ infected nodes, and $n_{\mathrm{R}}$ removed (either recovered or quarantined) nodes. Referring back to Fig.~\ref{fig:clique_states}, where a 3-node clique has four possible clique motifs and, for example, $Z_1 = (2,1,0)$ represents two susceptible nodes, one infected, and none recovered. Initially, a general clique with $c$ nodes will start with the clique motif $Z_1=(c-1,1,0)$. Given the probability of the susceptible nodes becoming infected, we can calculate the transition probability from $Z_1$ to either $Z_2$, $Z_3$, or inactive cliques with no new infected nodes. We can calculate the possible transition probability between all clique motifs in this fashion. Once we have the transition probabilities, it is easy to calculate the expected number of newly infected nodes from each clique motif transition and, therefore, calculate the mean matrix.

All this translates into the following general pipeline for processing any clique size we might wish, where we process the clique motifs one at a time with the following rules:

\begin{itemize}
    \item If the motif $Z_i$ has no infected nodes, we do nothing.
    \item If it has one or more infected nodes, we then create transitions to new motifs $Z_{j}=(n_{\mathrm{S}} + \delta n_{\mathrm{S}}, n_{\mathrm{I}}+ \delta n_{\mathrm{I}}, n_{\mathrm{R}}+ \delta n_{\mathrm{R}})$, for all $\delta n_{\mathrm{I}} \in \{ 1, \dots, n_{\mathrm{S}} \}$, $\delta n_{\mathrm{R}} \in \{ 0, \dots, n_{\mathrm{S}} - \delta n_{\mathrm{I}} \}$, and $\delta n_{\mathrm{S}} = -\delta n_{\mathrm{I}} - \delta n_{\mathrm{R}}$.
    \item If those new motifs have not been processed before, they are added to a queue for being processed. 
\end{itemize}
  
The transition probabilities $m_{ji}$ can then be computed for $j \neq 1$ by 
\begin{eqnarray}
\binom{n_{\mathrm{S}}}{\delta n_{\mathrm{I}}} \binom{n_{\mathrm{S}}- \delta n_{\mathrm{I}}}{\delta n_{\mathrm{R}}} p(n_{\mathrm{I}})^{\delta n_{\mathrm{I}}} (1-p(n_{\mathrm{I}}))^{n_{\mathrm{S}} +\delta n_{\mathrm{S}}} \nonumber \\  \cdot \alpha^{\delta n_{\mathrm{R}}}(1-\alpha)^{n_{\mathrm{S}}-\delta n_{\mathrm{R}}}
\,,
\end{eqnarray}
where the probability of $n_{\mathrm{I}}$ nodes causing a node to get infected is
\begin{equation}
p(n_{\mathrm{I}})=\sum_{k=0}^{n_{\mathrm{I}}-1} (\alpha -1 )^k (1-p)^k p\,.
\end{equation}

The element $m_{1i}$ can be computed by
\begin{equation}
    m_{1i}=(n_c-1)\sum_{j \neq 1} \delta n_{\mathrm{I}}(j,i) m_{ji}\,,
\end{equation}
where the value $\delta n_{\mathrm{I}}(j,i)$ is the value of  $\delta n_{\mathrm{I}}$ in the transition from $Z_i$ to $Z_j$.

This procedure will yield a sparse matrix, $\mathbf{M}$, where the size of the matrix and the number of non-zero elements grow as shown in Fig.~\ref{fig:Number of motifs}. This process allows us to automatically generate the mean matrices for any clique size we wish to examine efficiently. See \cite{simulations} for the Python implementation of this process.

\section{Impact of shortened quarantine time}\label{sec:shortQtime}
In the main text, we assumed that quarantine was forever, or effectively longer than the length of time a clique could have at least one infected node to propagate the infection. Here, we can consider this assumption's effect on the main three network topologies that we consider, where $c = 2,3,$ and $4$, where a node is only quarantined for a single timestep. This means a susceptible node placed in the quarantine compartment is returned to the susceptible compartment after one timestep.

\begin{figure}
     \includegraphics[width=\linewidth]{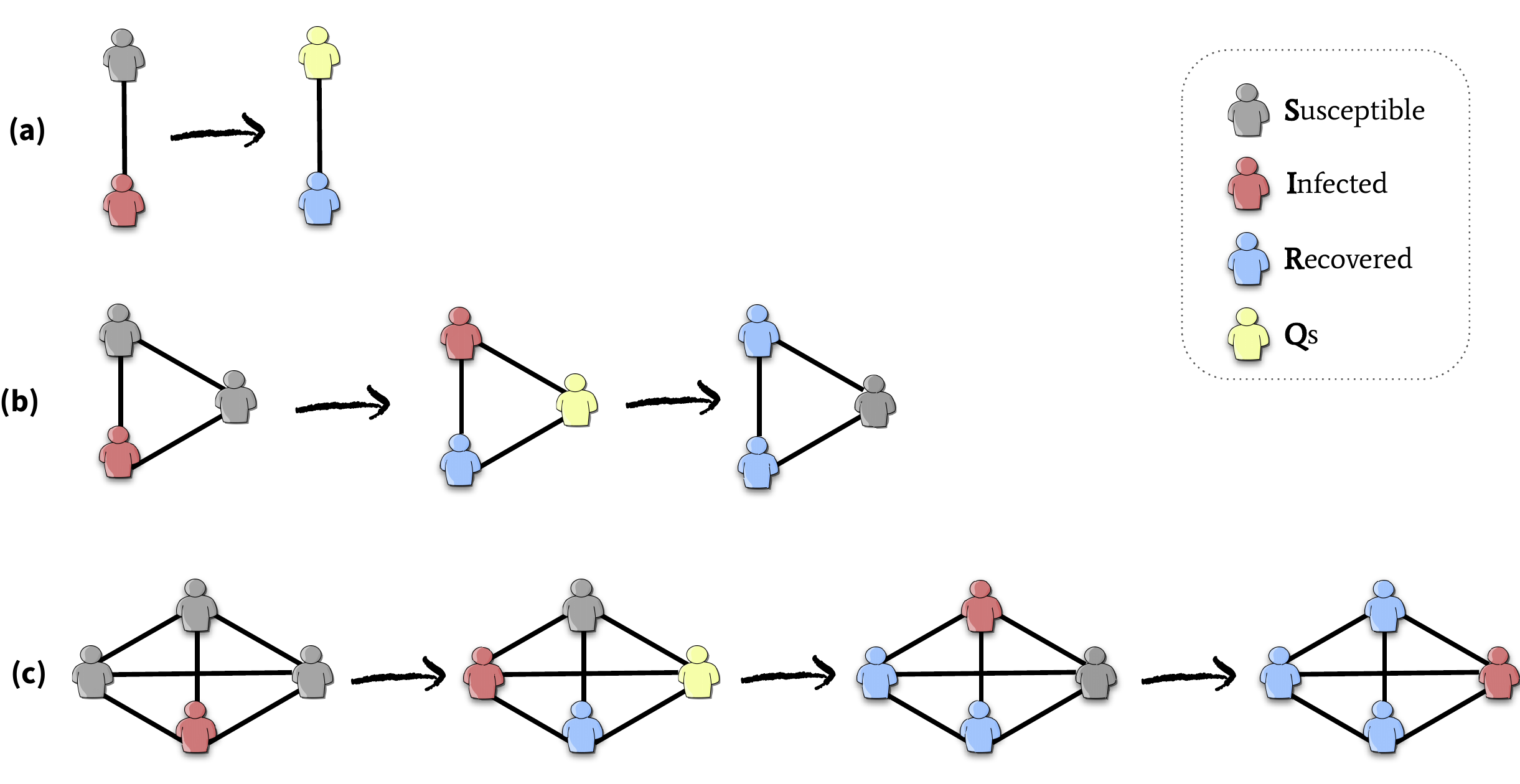}
    \caption{Schematic of the impact of the short-time quarantine. We track the number of susceptible, infected, and recovered nodes for the three main cliques we consider. we can note that the first instance where a previously quarantined susceptible node can meet an infected node appears in the $4$-clique.}
\label{fig:schshortQtime}
\end{figure}

For the $2$-clique case, see Fig.~\ref{fig:schshortQtime}a. This will not have any effect, as the clique only represents links, and any susceptible quarantined node will not have any infected neighbors in the following timestep. For the $3$-clique case, see Fig.~\ref{fig:schshortQtime}b. We have the same situation as the $2$-clique case. In any case, where the node is infected, and a susceptible node is in the quarantine compartment, in the following step, there will not be any active infected node to infect the returned susceptible node.

\begin{figure}
    \includegraphics[width=\linewidth]{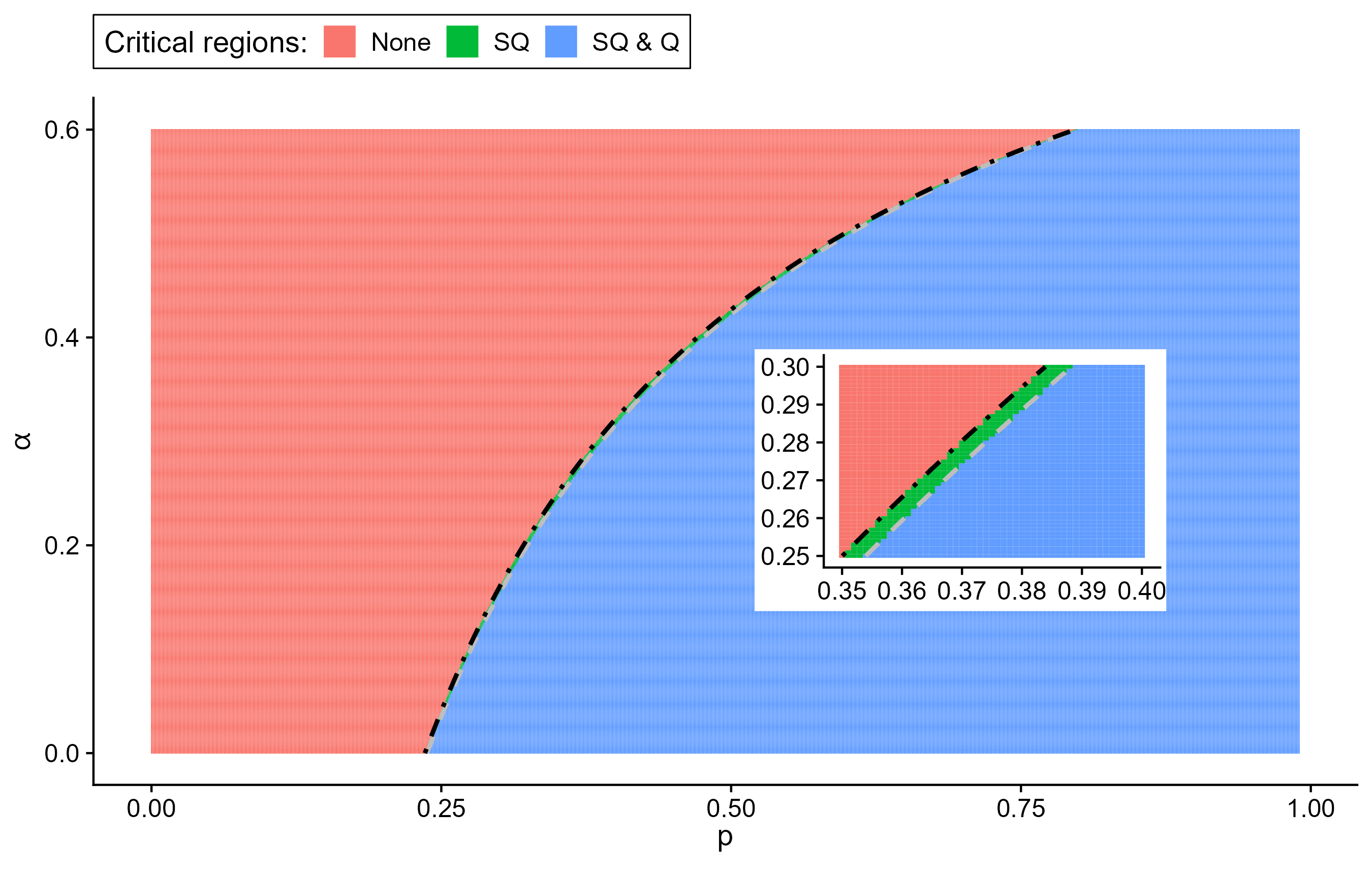}
    \caption{Differences in the critical regions obtained from the largest eigenvalues of the mean matrix for the original quarantine time (labeled Q) and the short quarantine time (labeled SQ). Inset shows a subset of parameter values from which the difference between the two quarantine times is more apparent.}
\label{fig:critialshortQtime}
\end{figure}

The $4$-clique is the only case where the shortened quarantine comes into play. This clique size has one possible infection path, illustrated in Fig.~\ref{fig:schshortQtime}c, that can result in an additional infection for this short quarantine time. If we were to readjust the mean matrix to account for this and recalculate the critical regions, see Fig.~\ref{fig:critialshortQtime}, we can see that it is hard to discern any difference in the overall behavior. If we concern ourselves with the inset of Fig.~\ref{fig:critialshortQtime}, which concentrates on a smaller parameter region, we can see that there is indeed a thin band of parameters for which the new quarantining behavior is super critical, but the original system is not. This difference is so small due to the presence of only a single set of low-probability events in which a returned susceptible node plays any role, and as such, our overall results are valid.

\begin{figure}
    \includegraphics[width=.7\linewidth]{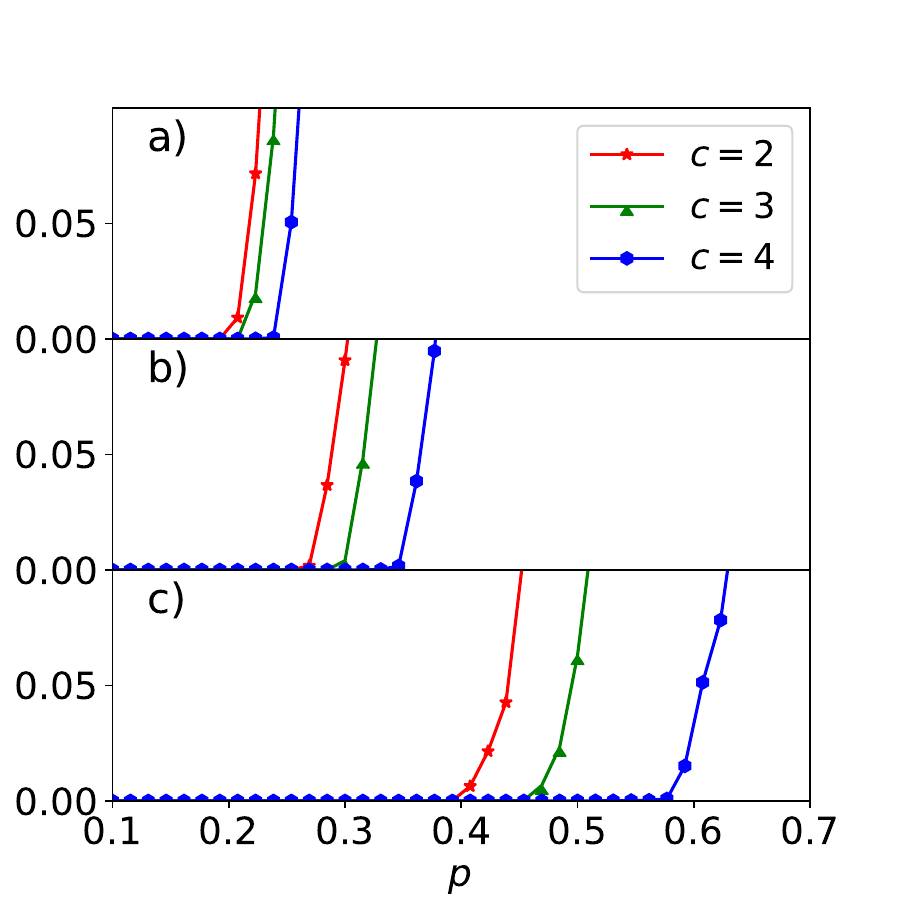}
    \caption{Phase transitions from a disease-free equilibrium to an endemic state for 2, 3, and 4-clique networks with degree 6 as introduced in Sec.~\ref{sec:Networks}. Nodes in quarantine can leave the $\mathrm{Q}$ compartment at every time step with probability $\alpha'= 0.4$. (a-c) The outbreak size $E$, normalized to the network size, is shown on the vertical axis for when (a) $\alpha = 0$ (no contact tracing), (b) $\alpha = 0.25$, and (c) $\alpha = 0.5$, from top to bottom respectively. Note that the transition points are shifted slightly to the right for larger clique sizes, $c$, even when there is no contact tracing ($\alpha=0$), but this difference is substantially amplified for larger $\alpha$ values. Results are based on Monte Carlo simulations introduced in Sec.~\ref{sec:SIR} and \cp{sec:shortQtime}.}
\label{fig:critialshortQtime2}
\end{figure}

To demonstrate that the exit of nodes from the quarantine $\mathrm{Q}$ compartment does not significantly alter the key epidemic outcomes we focus on, we have adjusted the model described in Section \ref{sec:SIR}. In this modified model, at each time step, nodes in quarantine have a probability \(\alpha'\) of leaving the Q compartment. Consequently, nodes from the $\mathrm{Q_S}$and $\mathrm{Q_I}$ compartments transition to the $\mathrm{S}$ and $\mathrm{R}$ compartments, respectively, with probability \(\alpha'\). This adjustment was made to reaffirm the results previously presented in Fig.~\ref{fig:phase-transition}a, specifically to show that these changes do not affect the epidemic thresholds of interest. Fig.~\ref{fig:critialshortQtime2} illustrates this scenario with \(\alpha' = 0.4\).

\subsection*{Next-generation Matrices for 4-cliques}
\label{sec:matrices}
\begin{figure}
    \includegraphics[width=\linewidth]{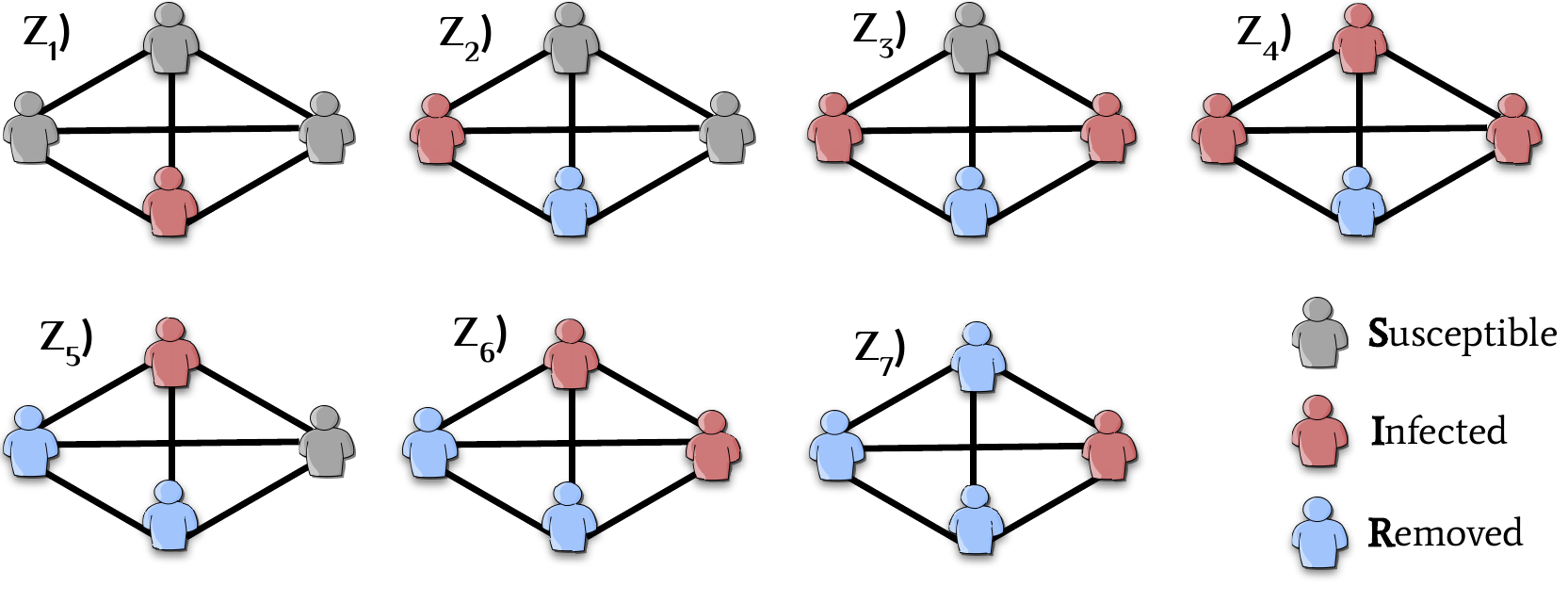}
    \caption{Life stages or diffusion patterns of a 4-clique. Similar to Fig.~\ref{fig:clique_states} for a 3-clique. We can generate the next-generation matrix for any clique size with our code introduced in Sec.~\ref{sec:auto_matrices}. }
\label{fig:clique_states2}
\end{figure}

Fig.~\ref{fig:clique_states2} shows the life stages of a 4-clique. For this case, the next-generation matrix according to the mean-filed and complex contagion approximations are given by Table~\ref{tab:m_matrix_4} and Table~\ref{tab:m_matrix_4_CC}, respectively.

\begin{table}
\begin{tabular}{ c c c c  } 
 \hline
 \hline
$i,j$ & $m_{ij}$ \\
\hline
$ 1 , 1 $&$ 3 p \left(- \alpha n_{c} + \alpha + n_{c} - 1\right) $\\
$ 1 , 2 $&$ 2 p \left(- \alpha n_{c} + \alpha + n_{c} - 1\right) $\\
$ 1 , 3 $&$ p \left(1-\alpha \right) \left(n_{c} - 1\right) \left(\left(\alpha - 1\right) \left(p - 1\right) + 1\right) $\\
$ 1 , 5 $&$ - p \left(\alpha - 1\right) \left(n_{c} - 1\right) $\\
$ 2 , 1 $&$ - 3 p \left(\alpha - 1\right)^{3} \left(p - 1\right)^{2} $\\
$ 3 , 1 $&$ 3 p^{2} \left(\alpha - 1\right)^{3} \left(p - 1\right) $\\
$ 4 , 1 $&$ - p^{3} \left(\alpha - 1\right)^{3} $\\
$ 5 , 1 $&$ - 6 \alpha p \left(\alpha - 1\right)^{2} \left(p - 1\right) $\\
$ 5 , 2 $&$ - 2 p \left(\alpha - 1\right)^{2} \left(p - 1\right) $\\
$ 6 , 1 $&$ 3 \alpha p^{2} \left(\alpha - 1\right)^{2} $\\
$ 6 , 2 $&$ p^{2} \left(\alpha - 1\right)^{2} $\\
$ 7 , 1 $&$ 3 \alpha^{2} p \left(1 - \alpha\right) $\\
$ 7 , 2 $&$ 2 \alpha p \left(1 - \alpha\right) $\\
$ 7 , 3 $&$ p \left(1-\alpha \right) \left(\left(\alpha - 1\right) \left(p - 1\right) + 1\right) $\\
$ 7 , 5 $&$ p \left(1 - \alpha\right) $\\
 \hline
 \hline
\end{tabular}
\caption{Non-zero elements of the next-generation matrix $\mathbf{M}$ for a 4-clique network as explained in Sec.~\ref{sec:m-field}. $m_{ij}$ gives the expected number of $Z_i$ cliques from a $Z_j$ clique, as shown in Fig.~\ref{fig:clique_states2}. $n_c$ is the number of cliques of size $c=4$ which for an $r$-regular $c$-clique satisfies the identity $n_c (c-1) = r$.}
\label{tab:m_matrix_4}
\end{table}

\begin{table}
\begin{tabular}{c c c c  } 
 \hline
 \hline
$i,j$ & $m_{ij}$ \\
\hline
\hline
$ 1 , 1 $&$ 3 p \left(n_{c} - 1\right) $\\
$ 1 , 2 $&$ - 2 \left(n_{c} - 1\right) \left(\left(\alpha - 1\right) \left(p - 1\right) - 1\right) $\\
$ 1 , 3 $&$ \left(n_{c} - 1\right) \left(\left(\alpha - 1\right)^{3} \left(p - 1\right)^{2} + 1\right) $\\
$ 1 , 5)$&$ \left(n_{c} - 1\right) \left(\left(\alpha - 1\right)^{2} \left(p - 1\right) + 1\right) $\\
$ 2 , 1 $&$ 3 p \left(p - 1\right)^{2} $\\
$ 3 , 1 $&$ 3 p^{2} \left(1 - p\right) $\\
$ 4 , 1 $&$ p^{3} $\\
$ 5 , 2$&$ 2 \left(1-\alpha \right) \left(p - 1\right) \left(\left(\alpha - 1\right) \left(p - 1\right) - 1\right) $\\
$ 6 , 2 $&$ \left(\alpha p - \alpha - p\right)^{2} $\\
$ 7 , 3 $&$ \left(\alpha - 1\right)^{3} \left(p - 1\right)^{2} + 1 $\\
$ 7 , 5 $&$ \left(\alpha - 1\right)^{2} \left(p - 1\right) + 1 $\\
\hline
\hline
\end{tabular}
\caption{Non-zero elements of the next-generation matrix $\mathbf{M}$ in the complex contagion approximation for a 4-clique network. $m_{ij}$ gives the expected number of $Z_i$ cliques from a $Z_j$ clique.}
\label{tab:m_matrix_4_CC}
\end{table}
\end{document}